\documentclass[%
 reprint,
superscriptaddress,
amsmath,amssymb,aps,
pra,notitlepage,longbibliography
]{revtex4-1}

\usepackage{natbib}
\usepackage{mathrsfs}
\usepackage{xr}
\usepackage{xcolor}
\usepackage{multirow}

\usepackage[...]{youngtab}

\usepackage{mathtools}
\DeclarePairedDelimiter\bra{\langle}{\rvert}
\DeclarePairedDelimiter\ket{\lvert}{\rangle}
\DeclarePairedDelimiterX\braket[2]{\langle}{\rangle}{#1 \delimsize\vert #2}

\newcommand{\fr}[2]{  \frac{#1}{#2} }           
\newcommand{\kb}[4]{  \ket*{#1, #2} \bra*{#3, #4}  }           
           
\newcommand{\tj}{\tilde{j} }

\begin{document}  

\title{Critical theory of non-Fermi liquid fixed point in multipolar Kondo problem}

\author{Adarsh S. Patri}
\affiliation{Department of Physics and Centre for Quantum Materials, University of Toronto, Toronto, Ontario M5S 1A7, Canada}

\author{Yong Baek Kim}
\affiliation{Department of Physics and Centre for Quantum Materials, University of Toronto, Toronto, Ontario M5S 1A7, Canada}

%\date{\today}

\begin{abstract}

When the ground state of a localized ion is a non-Kramers doublet, such localized ions may carry multipolar moments. 
For example, Pr$^{3+}$ ions in a cubic environment would possess quadrupolar and octupolar, but no magnetic dipole, moments. 
When such multipolar moments are placed in a metallic host, unusual interactions between these local moments and conduction electrons arise, in contrast to the familiar magnetic dipole interactions in the classic Kondo problem. 
In this work, we consider the interaction between a single quadrupolar-octupolar local moment and conduction electrons with $p$-orbital symmetry as a concrete model for the multipolar Kondo problem. 
We show that this model can be written most naturally in the spin-orbital entangled basis of conduction electrons. 
Using this basis, the perturbative renormalization group (RG) fixed points are readily identified. 
There are two kinds of fixed points, one for the two-channel Kondo and the other for a novel fixed point. 
We investigate the nature of the novel fixed point non-perturbatively using non-abelian bosonization, current algebra and conformal field theory approaches. 
It is shown that the novel fixed point leads to a, previously unidentified, non-Fermi liquid state with entangled spin and orbital degrees of freedom, which shows resistivity $\rho \sim T^{\Delta}$ and diverging specific heat coefficient $C/T \sim T^{-1 + 2\Delta}$ with $\Delta=1/5$.
Our results open up the possibility of myriads of non-Fermi liquid states, depending on the choices of multipolar moments and conduction electron orbitals, which would be relevant for many rare-earth metallic systems.
\end{abstract}

\maketitle
\section{Introduction}

Quantum theory of metals is often discussed in the framework of Fermi liquid,
where well-defined fermionic quasiparticles dominate thermodynamic and transport properties.
The discovery of metallic systems that demonstrated anomalous behaviors \cite{SSLee, Stewart}, such as diverging specific
heat coefficient and sub-quadratic temperature dependence of resistivity, prompted 
many-decades of research activities on non-Fermi liquid states. The number of concrete 
theoretical examples that are well understood is, however, not so large. 
The classic Kondo problem \cite{kondo_resistivity_1964,abrikosov_2_65,Anderson_poor_1970}, where the local magnetic dipole moment of a single 
localized ion interacts with conduction electrons' spin, 
is one prominent example where the dichotomy between Fermi liquid and non-Fermi liquid behaviours can be seen \cite{Hewson_1993, Coleman_perspective}.
We call this dipolar Kondo problem to clearly distinguish it from the model that we study
in this work. If we consider $m$-channels of conduction electrons interacting with the local
spin-1/2 moment, $m=1$ system is a Fermi liquid, where the local moment is screened
and becomes a part of Fermi sea \cite{nozieres_fl_picture, wilson_RG_1975, andrei_kondo_exact_single_1983, andrei_kondo_diag, affleck_current_algebra_cft}, while $m > 1$ systems lead to non-Fermi liquid states
with anomalous properties \cite{andrei_bethe_multi, bethe_multi_kondo, Tsvelick_bethe,    affleck_ludwig_overscreened,affleck_ludwig_fusion,affleck_two_channel_resistivity,affleck_cft_multi,affleck_ludwig_cft_bethe_compare, ludwig_exact_multi_review, overscreened_kondo_ybk, affleck_cft_review_1,affleck_cft_review_2}. This is an example where a single impurity can fundamentally
change the nature of the many-body ground state.

In this work, we demonstrate the existence of a novel non-Fermi liquid state 
in the multipolar Kondo problem, where the local moment is characterized by 
a non-Kramers crystal-field doublet that carries multipolar moments 
such as quadrupolar or octupolar moments.  
While such multipolar moments are abundant in quantum materials with $f$-electron moments \cite{multupole_rev_1, hidden_order_review, multupole_rev_3, rau_gingras_frustration, elasto_multipolar}, the corresponding Kondo problem \cite{cox_quad_kondo_1987, cox_zawadowski_book, cox_quad_kondo_hfm} has not been fully understood.
Taking the example of Pr$^{3+}$ ions in cubic environment, these ions support only quadrupolar and 
octupolar moments, and do not carry any magnetic dipole moment. 
If we introduce pseudo-spin-1/2 operators, ${\bf{S}} = (S^x, S^y, S^z)$ 
for the doublet, $S^x$ and $S^y$ ($S^z$) represent(s) quadrupolar (octupolar) moment.
While they satisfy the canonical SU(2) algebra, the physical contents of these
operators are very different from the spin-1/2 moment.
In particular, these multipolar moments do not couple solely to conduction electron spins,
but rather to conduction electron bilinears that transform in the same way that
the quadrupolar or octupolar moments transform.
In order to consider a concrete model, we take $p$-orbital bands
of conduction electrons, which belongs to the $T_2$ irreducible representation
of $T_d$ point group in cubic systems. 

We show that this model can be written most naturally if the spin-orbital
entangled basis, or the total angular momentum basis, is used for conduction electrons. 
This is interesting because the conduction electrons themselves do not
have any spin-orbit coupling in this model. 
It is the coupling to the multipolar moments which forces
the conduction electrons to have strong spin-orbital entanglement.
In this basis, it is shown that the perturbative RG fixed points can be
easily identified. As shown earlier in the non-entangled basis \cite{patri_kondo_2020}, there
exist two kinds of fixed points: one with behaviour of the two channel Kondo fixed point, and the other
being a novel fixed point. \textcolor{black}{In the perturbative RG analysis \cite{patri_kondo_2020, gan_coleman_andrei, Gan_1994}, the scaling dimension 
of the leading irrelevant operator at the novel fixed point is 1 + $\Delta$
with $\Delta=1/4$ \cite{patri_kondo_2020}, which leads to the resistivity $\rho \sim T^{1/4}$ and 
specific heat $C \sim T^{1/2}$ behaviours.} However, the nature of this novel fixed 
point was not clearly understood.
In addition, the stability of this fixed point beyond the perturbative analysis
was not addressed.
Since the novel fixed point may represent a previously unidentified non-Fermi liquid state, it is important to develop a deeper understanding of the nature of this fixed point.

Focusing on the novel fixed point, we first analyze the strong coupling limit
of the RG flow, \textcolor{black}{where the coupling constants are taken to be much larger than $\mathcal{O}(1)$}, and show that this strong coupling limit is unstable. 
This strongly suggests that the intermediate coupling fixed point found 
in the perturbative RG analysis is stable. 
In order to obtain non-perturbative results, we employ non-abelian 
bosonization, current algebra, and conformal field theory approaches to examine 
the critical theory of the novel fixed point.
Using the conformal embedding \cite{affleck_ludwig_fusion,cft_book}, the free theory of conduction electrons
can be written as a U(1) $\times$ SU(3)$_2$ $\times$ SU(2)$_3$ Kac-Moody invariant conformal field theory. The multipolar local moment
or the pseudo-spin-1/2 only couples to a sub-sector (three of 
eight generators) of SU(3)$_2$. These three generators form a closed algebra.
We highlight that this is not the 2-channel SU(3) Kondo model as only 
three generators are coupled to the multipolar local moment.
It is more useful to consider the coset construction SU(3)$_2$ = [3-state Potts model] $\times$ $\widetilde{\text{SU}}(2)_8$, where the multipolar
local moment then couples to the $\widetilde{\text{SU}}(2)_8$ sector.
Here $\widetilde{\text{SU}}(2)_8$ refers to SU(2)$_8$ with a convenient normalization of its generators.
{Considering the boundary conformal field theory \cite{cardy_bcft_1, cardy_bcft_2}, we find that 
the leading irrelevant operator at the novel 
fixed point is present in the $\widetilde{\text{SU}}(2)_8$ sector. 
If we consider a generalized model, U(1) $\times$ SU(3)$_k$ $\times$ SU($k$)$_3$,
this would correspond to an operator belonging to SU(2)$_{4k}$.}
The scaling dimension of {this operator is 1 + $\Delta$}
with $\Delta = 2 / (4k + 2)$. The perturbative fixed point corresponds
to the large $k$ limit, and hence $\Delta = 1/ (2k)$. With $k=2$ (as in our case), this reproduces
1/4 in the perturbative RG analysis. 
The corresponding exact scaling dimension is $\Delta = 1/5$ for $k=2$.
This leads to singular behaviour for experimentally relevant quantities, such as the specific heat 
coefficient $C/T \sim T^{-1 + 2\Delta} = T^{-3/5}$ and
the resistivity $\rho \sim T^{\Delta} = T^{1/5}$.
\textcolor{black}{This represents a rare-example of solvable non-Fermi liquid fixed points. 
In the broader context, our work provides a concrete example of the possibility of a wide variety of Kondo effects, as well as a myriad of non-Fermi liquids, that may arise in rare-earth metallic compounds.}

\section{Microscopic Model}

The combination of spin-orbit (SO) coupling and crystalline electric fields (CEFs) in rare-earth compounds allows for the development of exotic higher-rank multipolar moments.
Taking a localized Pr$^{3+}$ ion in a cubic environment as a concrete example, the SO-coupled $J=4$ multiplet of $f^2$ electrons is split by the CEF to give rise to a low-lying non-Kramers $\Gamma_{3g}$ doublet \cite{review_exotic_multipolar}.
This $\Gamma_{3g}$ doublet can support time-reversal even quadrupolar moments $\mathcal{O}_{20} = \frac{1}{2} (3J_z^2 - J^2)$, $\mathcal{O}_{22} = \frac{\sqrt{3}}{2} (J_x^2 - J_y^2)$, as well as a time-reversal odd octupolar moment $\mathcal{T}_{xyz} = \frac{\sqrt{15}}{6} \overline{J_x J_y J_z}$, where the overline represents a fully symmetrized product.
These multipolar moments can be efficiently described by the pseudospin-1/2 operator ${\bf{S}} = \left( S^x, S^y, S^z \right)$,
%%%%%%
\begin{align}
&S^x = -\frac{1}{4}\mathcal{O}_{22},  ~~~ ~~S^y = -\frac{1}{4}\mathcal{O}_{20},  ~~~ ~~S^z = \frac{1}{3\sqrt{5}}\mathcal{T}_{xyz}. 
\end{align}
%%%%%%
Embedding such multipolar moments in a metallic system, the localized electronic configuration can fluctuate from its $f^2$ ground state configuration ($\Gamma_{3g}$) to excited $f^1$ states ($\Gamma_7$) via hybridization with the sea of conduction electrons.
Group theoretically, this hybridization process occurs only if the conduction electrons possess the appropriate symmetry i.e. $\Gamma_c = \Gamma_{3g} \otimes \Gamma_{7} = \Gamma_8$, where $\Gamma_c$ denotes the irrep of the conduction electron states \cite{cox_zawadowski_book}.
As described in a recent work \cite{patri_kondo_2020}, one way to form this $\Gamma_8$ irrep is from the combination of cubic $p$-like orbitals equipped with spinor-1/2 degree of freedom, $\Gamma_c = p \otimes \frac{1}{2} = \Gamma_8 \oplus \Gamma_6$.

The natural, physical setting for such a construction is in the family of cubic rare-earth compounds, Pr(Ti,V)$_2$Al$_{20}$ (PrIr$_2$Zn$_{20}$), where the Pr$^{3+}$ ions are subjected to a local $T_d$ symmetry by a surrounding cage of Al (Zn) atoms \cite{hfm_superconductivity_v, hattori_afq_fq_2014, sbl_ybk_landau_2018, patri_unveil_2019, pr_nfl, quad_nfl_onimaru, ming_quadrupolar_kondo}.
Focussing on the choice of conduction $p$ electron orbitals, the $T_d$ symmetry-permitting couplings of the conduction electrons to a local moment (located at impurity site $\bf{x}=0$) are,
%%%
\begin{align}
H_{1} &= K_{1} \left[ S^x \left( c_{x,\alpha}^{\dagger} c_{x,\beta} - c_{y,\alpha}^{\dagger} c_{y,\beta} \right) \right. \label{h_q_1} \\ 
& \left. +  \frac{S^y}{\sqrt{3}} \left( 2c_{z,\alpha}^{\dagger} c_{z,\beta} - c_{x,\alpha}^{\dagger} c_{x,\beta} - c_{y,\alpha}^{\dagger} c_{y,\beta} \right)  \right] \delta_{\alpha \beta} \nonumber
\end{align}
%%%%%
\begin{align}
{H}_{2} = &-\sqrt{3}  K_{2}  \left[  S^x \left( i \sigma^x_{\alpha \beta} c_{y,\alpha}^{\dagger} c_{z,\beta}  -  i \sigma^y_{\alpha \beta}  c_{z,\alpha}^{\dagger} c_{x,\beta} \right)  \right. \label{h_q_2}  \\ 
& \left. +  \frac{S^y}{\sqrt{3}} \left( 2 i \sigma^z_{\alpha \beta}  c_{x,\alpha}^{\dagger} c_{y,\beta} - i \sigma^x_{\alpha \beta} c_{y,\alpha}^{\dagger} c_{z,\beta}   - i \sigma^y_{\alpha \beta} c_{z,\alpha}^{\dagger} c_{x,\beta}  \right) + \text{h.c.} \right] \nonumber
\end{align}
%%%%%  
\begin{align}
H_3 = K_3  S^z  \left[ \sigma^x_{\alpha \beta} \left( c_{y,\alpha}^{\dagger} c_{z,\beta} + \text{h.c.} \right) + \text{cyclic.} \right]
\label{h_o_1}
\end{align}
%%%
where $c_{\mathcal{P},\alpha}$ denotes the conduction electron annihilation operator at site-0, orbital $\mathcal{P}$ and spin $\alpha$, and an implicit summation over conduction spin-indices $\alpha, \beta = \{ \uparrow, \downarrow \}$.
We note that this construction is based on symmetries, and hence is broadly applicable to cubic \textcolor{black}{($T_d$)} systems.
We note that in terms of cubic harmonics these terms can be easily be seen as satisfying the $T_d$ symmetry.
In particular, we note that since $S^x$ has a $x^2 - y^2$ symmetry, it respectively couples to $x^2 - y^2$ charge densities in Eq. \ref{h_q_1} and currents in Eq. \ref{h_q_2}.
We stress that in the $T_d$ point group $x$ and $yz$ transform identically, and correspondingly for the cyclic permutations.
Thus, Eq. \ref{h_q_1} and \ref{h_q_2} have the same symmetry structure.

As studied in Ref. \onlinecite{patri_kondo_2020}, performing third order perturbative renormalization group calculations leads to two non-trivial fixed points (each of these come as a pair, which are related by a canonical transformation of the pseudospin)
(i) Fixed point I: $K_1 = - \sqrt{3} K_2 = K_3 = \frac{1}{2\sqrt{3}}$, and (ii) Fixed point II: $K_1 = \sqrt{12}K_2 =  -\frac{1}{2\sqrt{6}}$, $K_3 = -\frac{1}{4\sqrt{3}}$.
Fixed point I was found to have the same exponents for physical properties from perturbative RG as the two-channel Kondo model, \textcolor{black}{while fixed point II possessed highly singular scaling characterized by a leading irrelevant operator of dimension $1+\Delta$, where $\Delta=1/4$ is the slope of $\beta$-function at the fixed point.}
We call this the novel fixed point.
The perturbative scaling can be easily understood by rewriting the beta-function from Ref. \onlinecite{patri_kondo_2020} in terms of a single coupling constant, $g_k$, by fixing the ratios of the original $K_{1,2,3}$ couplings as that at the fixed point of interest.
For fixed point II, we can define $K_1 = - \frac{1}{2 \sqrt{6}} g_k$, $K_2 = - \frac{1}{12 \sqrt{2}} g_k$, $K_3 = -\frac{1}{4\sqrt{3}} g_k$, such that when $g_k \rightarrow 1$, we arrive at the fixed point II.
This leads to the $\beta$-function,
%%%
\begin{align}
\frac{d g_k}{d \ln D} = -\frac{g_k^2}{4} + \frac{g_k^3}{4}
\end{align}
%%%
where $D$ is the UV cutoff.
The fixed point is located at $g_k ^*= 1$, and the slope of the beta-function at the fixed point is $\Delta = 1/4$.
Indeed this $\Delta$ appears in the leading specific heat and resistivity scaling exponents, as was found in the original RG calculation in Ref. \onlinecite{patri_kondo_2020}.

\section{Spin-orbit coupled basis for conduction electrons}
\label{sec_so_coupling_basis}
\begin{figure}[t]
\includegraphics[width=0.975\linewidth]{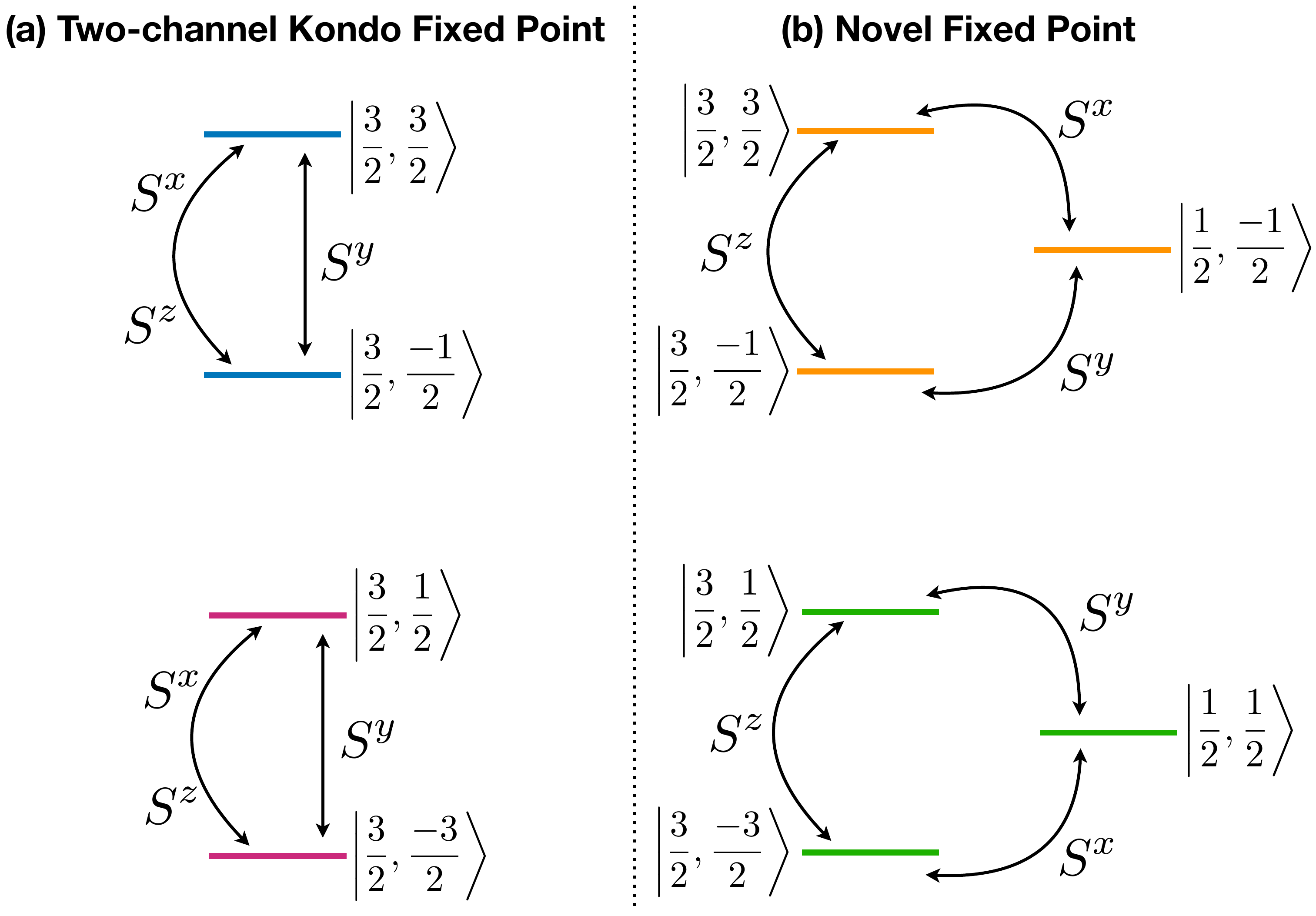}
\caption{Schematic picture of Kondo Hamiltonians tuned to each of the fixed points, with conduction electron transitions amongst different $\ket{j, m_j}$ states via coupling to the multipolar moments.
(a) Two-channel Kondo fixed point, involving two decoupled pairs of conduction electrons from $j=3/2$ sector (blue and purple). 
(b) Novel fixed point, involving two decoupled triplet of states from $j=1/2$ and $j=3/2$ conduction electron levels (orange and green).}
\label{fig_kondo_hamiltonians}
\end{figure}
Though the cubic harmonics enables the symmetry nature of the coupling to be easily verified, it does not give immediate indication as to the underlying nature of the fixed points.
In order to shed light on this, we consider double change of basis: (i) from cubic harmonics to spherical harmonics, and then to (ii) spin-orbit coupled basis by implementation of Clebsch-Gordon angular momentum addition.
The change of bases are delineated in Appendix \ref{app_change_of_basis}.
The above Kondo couplings can be recaptured into the form, 
%%%
\begin{widetext}
\begin{align}
\label{eq_ham_j_total}
H_{\text{tot}} &= \left( - \frac{K_{1}}{\sqrt{3}} + 2 K_{2} \right) \Bigg[ S^x \left( \kb{\frac{3}{2}}{ \frac{1}{2} }{\frac{3}{2}}{\frac{-3}{2}} + \kb{\frac{3}{2}}{ \frac{3}{2} }{\frac{3}{2}}{\frac{-1}{2}}  + \text{h.c.}    \right) \Bigg. \nonumber \\
\Bigg. &~~~~~~~~~~~ ~~~~~~~~~ ~~~~~~ ~ + S^y \left(  \kb{\frac{3}{2}}{ \frac{-3}{2} }{\frac{3}{2}}{\frac{-3}{2}} + \kb{\frac{3}{2}}{ \frac{3}{2} }{\frac{3}{2}}{\frac{3}{2}} - \kb{\frac{3}{2}}{ \frac{1}{2} }{\frac{3}{2}}{\frac{1}{2}}  - \kb{\frac{3}{2}}{ \frac{-1}{2} }{\frac{3}{2}}{\frac{-1}{2}}       \right) \Bigg] \nonumber \\ 
%%%
&+ \left(  \frac{\sqrt{2}K_{1}}{\sqrt{3}} + \sqrt{2} K_{2} \right) \Bigg[ S^x \left( \kb{\frac{3}{2}}{ \frac{3}{2} }{\frac{1}{2}}{\frac{-1}{2}} - \kb{\frac{3}{2}}{ \frac{-3}{2} }{\frac{1}{2}}{\frac{1}{2}}      \right) 
+ S^y \left(  \kb{\frac{3}{2}}{ \frac{-1}{2} }{\frac{1}{2}}{\frac{-1}{2}} - \kb{\frac{3}{2}}{ \frac{1}{2} }{\frac{1}{2}}{\frac{1}{2}}     \right)  + \text{h.c.}   \Bigg] \nonumber \\
& +  \left( \sqrt{3} K_{3} \right)S^z \Bigg[  i	 \kb{\frac{3}{2}}{ \frac{-3}{2} }{\frac{3}{2}}{\frac{1}{2}}  + i  \kb{\frac{3}{2}}{ \frac{3}{2} }{\frac{3}{2}}{\frac{-1}{2}} + \text{h.c.}	\Bigg]
\end{align}
\end{widetext}
%%%
where we use ket (bra) notation of $\ket{j, m_j}$ ($\bra{j', m_{j'}}$) to denote conduction electron creation $c_{j,m_j} ^{\dagger}$ (annihilation $c_{j',m_{j'}}$) operators of total angular momentum $j$ ($j'$) and and $z$-projection $m_j$ ($m_{j'}$); the impurity site-location ($\bf{x}=0$) is dropped for brevity.
Equation \ref{eq_ham_j_total} sheds a remarkable insight into the nature of the perturbative fixed points.

\subsection{Two-channel Kondo fixed point}
First, we consider tuning of the coupling constants to fixed point I.
The second term in Eq. \ref{eq_ham_j_total} vanishes and the remaining collection of coupling constants become $\left( - \frac{K_{1}}{\sqrt{3}} + 2 K_{2} \right) \rightarrow 1/2$ and 
$\left( \sqrt{3} K_3 \right) \rightarrow 1/2$.
The remaining terms only involve the four $j=3/2$ states, which decouple to into two independent (time-reversal related) doublets.
Defining pseudospin-1/2 operators ${\bf{\tau}}^{x,y,z}_A$ and $\tau^{x,y,z}_B$ for each of the decoupled doublets (Fig. \ref{fig_kondo_hamiltonians}(a): blue and purple levels, respectively), 
Eq. \ref{eq_ham_j_total} can be rewritten as,
%%%
\begin{align}
\hspace{-2mm} H_{\text{tot}}=  S^x \left(\tau_A ^x + \tau_B^x \right) - S^y \left( \tau_A ^z + \tau_B^z \right) + S^z \left( \tau_A ^y + \tau_B^y \right),
\end{align}
%%%
where we present the form of the ${\vec{\tau}}_{A,B}$ operators in Appendix \ref{app_basis_two_channel}.
This is precisely the form of the conventional two-channel Kondo model at its fixed point, and thus confirms our perturbative determination that fixed point I has two-channel non-Fermi liquid behaviour. 

\subsection{Novel fixed point}
\label{sec_novel_fp}

Tuning the coupling constants to the novel fixed point (II), the first term in Eq. \ref{eq_ham_j_total} vanishes and the remaining collection of coupling constants become $\left(  \frac{\sqrt{2}K_{1}}{\sqrt{3}} + \sqrt{2} K_{2} \right) \rightarrow -1/4$ and $\left( \sqrt{3} K_3 \right) \rightarrow -1/4$.
Interestingly, the remaining terms do not belong to a single $j$ manifold, but involve terms from both $j=3/2$ and $j=1/2$.
This is unlike the above two-channel model, which only involved conduction $j=3/2$ states.
The terms can be organized into two decoupled triplet of states:  and (i) $\left\{  \ket*{\fr{3}{2}, \fr{3}{2}}, \ket*{\fr{3}{2}, \fr{-1}{2}}, \ket*{\fr{1}{2}, \fr{-1}{2}}	\right\}$ and (ii) $\left\{  \ket*{\fr{3}{2}, \fr{1}{2}}, \ket*{\fr{3}{2}, \fr{-3}{2}}, \ket*{\fr{1}{2}, \fr{1}{2}}	\right\}$, where we again use the notation of $\ket{j, m_j}$. Performing another unitary rotation about the $\ket*{\fr{1}{2}, \fr{-1}{2}}$ axis in the (ii) space, reduces the Kondo coupling into the following elegant form,
%%%
\begin{align}
\hspace{-2mm}  H_k = \sum_{m=1,2} \vec{\psi}_m ^{\dag} (0) \left[ \frac{S^x}{2}   \frac{\lambda^4}{2}  + \frac{S^y}{2} \fr{\lambda^6}{2}  + \frac{S^z}{2} \fr{\lambda^2}{2}  \right] \vec{\psi}_m (0)
\label{eq_kondo_decoupled_novel}
\end{align}
%%%
where $\lambda^a$ are the SU(3) Gell-Mann matrices, $m$ labels the two decoupled bases i.e. (i) $\vec{\psi}_{m=1} ^\dag = \left( - c_{\frac{3}{2}, \frac{3}{2}}^{\dag}, - c_{\frac{3}{2}, \frac{-1}{2}}^{\dag}, c_{\frac{1}{2}, \frac{-1}{2}}^{\dag} \right)$ and (ii) $\vec{\psi}_{m=2}^\dag = \left( c_{\frac{3}{2}, -\frac{3}{2}}^{\dag}, c_{\frac{3}{2}, \frac{1}{2}}^{\dag}, c_{\frac{1}{2}, \frac{1}{2}}^{\dag} \right)$, where the negative signs indicate the aforementioned final unitary transformation.
Equation \ref{eq_kondo_decoupled_novel} can be schematically visualized in Fig. \ref{fig_kondo_hamiltonians}(b).
Before examining the nature of this fixed point in detail, we first consider the justification of its existence from a strong-coupling limit analysis.

\section{Instability of Strong \\ Coupling Limit}

The strong-coupling limit provides a consistency rationale for the existence of the perturbatively obtained fixed point.
In the strong-coupling limit, the kinetic term can be ignored, and the problem reduces to determining the degeneracy of the ground state at the impurity site.
If the ground state is non-degenerate, this indicates the local moment has been quenched and no further Kondo effects can occur.
In the original Kondo model, this is the outcome, which is in agreement with the picture of a formation of a Kondo singlet.
If, on the other hand, the ground state is degenerate, this indicates that even at strong coupling there is still enough `freedom' at the impurity site to participate in further Kondo scattering events with conduction electrons. 
This degeneracy thus indicates the instability of the strongly coupled fixed point, and indicates a renormalization group flow away from it and towards an intermediate fixed point \cite{noz_fixed_point}.

Focussing on the novel fixed point (II), the strong coupling limit results in a four-fold degenerate ground state (Appendix \ref{app_strong_coupling}).
This four-fold ground state degeneracy is larger than the two-fold degeneracy one obtains from the two-channel Kondo model, which underlines the increased quantum nature of the novel fixed point.
Indeed, even with the introduction of perturbations to the strongly-coupled Hamiltonian (such as conduction tunnelling processes to/from the impurity), the ground state retains (at least partially) its degeneracy.
Thus, even with perturbations the strongly coupled fixed point retains sufficient degrees of freedom to scatter with conduction electrons and drive the system away from the strongly coupled fixed point, and towards an intermediate fixed point.

\section{Current Algebra Approach to novel fixed point}

In the same spirit as the original Kondo problem (\textcolor{black}{and even rare-earth impurity systems \cite{cox_zawadowski_book}}), we consider radial (s-wave) scattering events, which allows the three-dimensional free-fermion model to be mapped to a chiral one-dimensional problem \cite{affleck_ludwig_overscreened}, 
%%%
\begin{align}
H_{0} = \int_{-\infty}^{\infty} dx \sum_{\beta=1,2,3; t=1,2}  \psi_{L, \beta, t}^{\dag} (x) (i \partial_x ) \psi_{L, \beta, t} (x)
\label{eq_free_fermion}
\end{align}
%%%
where $x$ is the radial coordinate, $\psi_{L,\beta,j}$ denotes a left-moving fermionic field with SU(3) triplet-space label $\beta$, $t$ is the channel index, $\beta$ sums over the respective triplet subspace (i) and (ii) described in Sec. \ref{sec_novel_fp}, and $v_F$ the Fermi velocity is set to 1.
\textcolor{black}{The identical Fermi velocity in Eq. \ref{eq_free_fermion} for all the six flavours of fermions indicates a SU($3\times2$) symmetry, which arises from the exact microscopic symmetries of the lattice.
One can understand this by considering the conduction electron kinetic term in the original cubic harmonic ($p$ orbitals) basis of Eq. \ref{h_q_1}, \ref{h_q_2}, \ref{h_o_1}, and focussing on the Fermi surface that is well localized about the zone-centre \cite{dhva_nagashima_2013, dhva_nagashima_2020}.
In particular, a three-fold degeneracy of the $p_x, p_y, p_z$ orbitals satisfies the underlying cubic ($O_h$) symmetry about this high-symmetry point, and a two-fold spinful (up/down) degeneracy arises as a natural consequence of the lack of spin-orbit coupling for the (time-reversal invariant) bare conduction electrons.
Translating to the spin-orbit coupled basis of Sec. \ref{sec_so_coupling_basis} yields Eq. \ref{eq_free_fermion} with the SU(6) symmetry.}

For definiteness, we consider the left-moving electrons as living on a circle of length $2l$.
\textcolor{black}{Employing standard quantum mechanics,} with the antiperiodic boundary conditions $\psi_L(-l) = - \psi_L(l)$, this leads to (in momentum space) the allowed fermionic wavevectors to be $k = \frac{\pi}{l} (n + 1/2)$, where $n \in \mathbb{Z} \geq 0$.
\textcolor{black}{Alternatively, a (1+1)-dimensional free fermion system can be described in terms of a conformal field theory.
As well, the (spatial) one-dimensionality of the problem encourages one to bosonize the fermionic theory in terms of mutually commuting bosonic current operators.
There are a number of different ways by which the different degrees of freedom of the fermions can be partitioned, while preserving the conformal invariance, in a procedure known as conformal embedding.
The choice of the partitioning, as will become clear, depends on the problem one is considering. 
In any of the embeddings, only certain combinations of the states associated with the decoupled bosonic current operators can be combined so as to recover the original fermionic excitation spectrum; these combinations are called ``gluing'' conditions \cite{affleck_ludwig_fusion}.
}
\textcolor{black}{
One way to decompose the above free-fermion theory of 2$\times3=6$ different flavours of fermions is in terms of its U(1) charge and SU(6) flavour degree of freedom.
Here each fermion flavour carries its own U(1) charge, $Q_i$, such that the total charge $Q = \sum_{i=1}^6 Q_i$.
The conformal embedding of rewriting Eq. \ref{eq_free_fermion} in terms of U(1) and SU(6)$_1$ bosonic currents has the ``gluing'' condition of $Q = p$(mod 6), where $p$ is the highest state of SU(6)$_1$ \cite{affleck_ludwig_fusion}.
We note that the subscript denotes the level of the Kac-Moody (KM) algebra.}
\textcolor{black}{However, the kinetic term and Kondo coupling suggest another conformal embedding:} in terms independent U(1) charge, SU(3) flavour and SU(2) spin degrees of freedom.
We define the following left-moving non-Abelian currents in terms of the complex space-time variable $z \equiv \tau + i x$, 
%%%
\begin{align}
J (z) &= \sum_{\substack{\beta=\{1,2,3\} \\ t=\{1,2\}} } : \psi_{L,\beta, t} ^{\dag} \psi_{L,\beta, t} : (z) \label{eq_charge_current}  \\
J ^a(z) &=  \sum_{\substack{\alpha,\beta=\{1,2,3\} \\ t=\{1,2\}}}: \psi_{L,\alpha, t} ^{\dag} \left( \frac{ \lambda_{\alpha \beta} ^{a}}{2}  \right) \psi_{L,\beta, t} :  (z) \label{eq_gellmann_current}  \\
J ^A(z) &=  \sum_{\substack{\beta=\{1,2,3\} \\ t,q=\{1,2\}}}  : \psi_{L,\beta, t} ^{\dag} \left( \frac{ \sigma_{t q} ^{A}}{2}  \right) \psi_{L,\beta, q} : (z) \label{eq_spin_current}
\end{align}
%%%
where the three currents in Eq. \ref{eq_charge_current}, \ref{eq_gellmann_current}, \ref{eq_spin_current} denote the U(1) charge, SU(3) Gell-Mann, and SU(2) spin bosonic currents, respectively, $a=\{1,2,...8\}$ sums over the eight SU(3) (Gell-Mann) generators and $A=\{x,y,z\}$ sums over the three SU(2) (Pauli) generators, and``: \dots \ :'' refers to normal ordering by point splitting i.e. $:\psi^{\dag} \psi:(z) \equiv \lim_{\delta \rightarrow 0} \left[ \psi^{\dag}(z-i\delta)\psi (z) - \langle \psi^{\dag}(z-i\delta)\psi (z) \rangle \right]$.
In effect, bosonizing the theory with these non-abelian currents amounts to the decomposition of the irreps of the SU(6)$_1$ in terms of 
irreps of SU(3)$_2 \times$SU(2)$_3$.
\textcolor{black}{This decomposition preserves the central charge (a sufficient condition for preserving conformal invariance \cite{affleck_ludwig_fusion, cft_book})}, and importantly allows a relationship to be established between the U(1) charge ($Q$) and the irrep labels for SU(3)$_2$ and SU(2)$_3$.
We present the conformal branching rules of SU(3)$_2 \oplus $ SU(2)$_3 \subset$ SU(6)$_1$ in Appendix \ref{app_conformal_embedding}.

%\textcolor{black}{conformal embedding!}

These bosonic currents satisfy the following respective KM Algebra, which can be obtained (Appendix \ref{app_opes}) from the mode expansions of the operator product expansions (OPEs),
%%%
\begin{align}
& \left[J_{n}, J_{m} \right] = 6 n \delta_{n+m} \\
& \left[ J_{n}^a, J_{m}^b \right] =  i f^{abc} J_{n+m} ^c + \frac{2}{2} n \delta^{ab} \delta_{n+m} \\
& \left[ J_{n}^A, J_{m}^B \right] =  i \epsilon^{ABC} J_{n+m} ^C + \frac{3}{2} n \delta^{AB} \delta_{n+m}
\end{align}
%%%
where we introduce the modes from the Laurent expansion $J^{a} (z) = \sum_{n \in \mathbb{Z}} z^{-n-1}J^{a}_{n}$ etc., and $f^{abc}$ and $\epsilon^{ABC}$ are the structure constants of the SU(3) and SU(2) Lie algebras respectively.
We note that the normalization of the highest root is set to 1 (the canonical convention) to compute the above KM algebra.
As seen, the SU(3) Gell-Mann and SU(2) spin currents satisfy SU(3)$_2$ and SU(2)$_3$ KM algebra, respectively.
Using these bosonic currents, the kinetic term can thus be recaptured in the following Sugawara form,
%%%
\begin{align}
\hspace{-3.4mm} H_0 = \frac{1}{12} : J J : (z) + \frac{1}{5} : J ^a J ^a : (z) + \frac{1}{5} :J^A J ^A : (z) .
\label{eq_sugawara_free}
\end{align}
%%%
The free-fermion theory is a U(1)$\times$SU(3)$_2 \times$SU(2)$_3$ KM invariant conformal field theory.
The Kondo coupling of Eq. \ref{eq_kondo_decoupled_novel} can also be rewritten in terms of these bosonic currents,
%%%
\begin{align}
H_K = g_k \left( J^4 (z) \frac{S^x}{2} + J^6 (z) \frac{S^y}{2} + J^2 (z) \frac{S^z}{2} \right)
\label{eq_select_generators}
\end{align}
%%%
where we tune the ratio of the couplings such that the perturbative fixed point is reached at $g_k^* = 1$.

The Sugawara form suggests that we may `complete the square' and absorb the Kondo coupling into the free Hamiltonian (with the addition of a trivial energy constant) by redefining a `shifted' bosonic current. 
Of course, this can only be done for certain special values of the Kondo coupling constant, $g_k$.
In our case, when $g_k ^c =  2/5$, $S^{x,y,z} / 2$ can be `absorbed' into their respective $J^{4,6,2}$ currents to give the Sugawara form,
%%%
\begin{align}
H_0 + H_K = & \frac{1}{12} : {J} {J} : (z) + \frac{1}{5} : \mathcal{J} ^a \mathcal{J} ^a : (z) \nonumber \\
& + \frac{1}{5} : {J} ^A {J} ^A : (z) ,
\label{eq_sugawara_complete}
\end{align}
%%%
where $\mathcal{J} ^4 \equiv J^4 + \frac{S^x}{2}$, $\mathcal{J} ^6 \equiv J^6 + \frac{S^y}{2}$, $\mathcal{J} ^2 \equiv J^2 + \frac{S^z}{2}$, and the remaining currents are unaffected i.e. $\mathcal{J} ^a = J^a$ for $a \neq \{2,4,6\}$ etc.
Importantly, these `absorbed' SU(3) Gell-Mann currents satisfy the same SU(3)$_2$ KM algebra.
In the same spirit as Affleck and Ludwig \cite{affleck_cft_review_1}, we interpret Eq. \ref{eq_sugawara_complete} as the effective Hamiltonian and $g_k ^c = 2 / (k+3) = 2/5$ as the coupling constant at the infrared (IR) fixed point.  
In that sense, both UV ($g_k = 0$) and IR ($g_k ^c = 2/5$) have the same U(1)$\times$SU(3)$_2$$\times$SU(2)$_3$ symmetry.
Finally, we notice that the IR coupling constant $g_k ^c$ reduces to the the perturbative fixed point $g_k ^*$ in the limit of large $k$ and then setting $k\rightarrow 2$.

\textcolor{black}{
Indeed, the `absorption' of the multipolar impurity into the free electron theory is a consequence of assuming that the critical point corresponds to a conformally invariant boundary condition on the bulk conformal theory.
The reasonableness of this assumption can be understood by considering the correlation function between two points far away from the impurity \cite{affleck_cft_review_1}.
The impurity, we recall, is located at the origin, and so we can be visualize it as the space-(imaginary)time boundary above which the free-fermions and Kondo interaction exists.
Now, since the points are far from the boundary we naturally expect the correlation function to probe the bulk free-fermion (conformal) behaviour. %, without any influence from the bounadry.
However, if the two points are closer to the boundary than to each other, it is now prudent to expect that the boundary can influence the correlation function. % (as we are examining correlations on length scales that include the distance to the boundary).
As such, the impurity acts as a termination of the bulk conformal theory behaviour i.e. as a boundary condition to the bulk theory.
Indeed, since the impurity is fixed at $\bf{x}=0$, under the conformal transformation, we thus expect the impurity to act as a \textit{conformally invariant} boundary condition to the bulk theory.
In the renormalization group sense, the impurity is absorbed into the continuum fermion theory for increasing length scales, and its influence is retained as a modification of the free-fermion boundary condition.
}

Interestingly, only select SU(3) generators have absorbed the SU(2) impurity, and one may notice that the absorbed impurity comes with a factor of $1/2$.
This $1/2$ factor is the crucial difference in this model, as it relates the \{2-4-6\} subalgebra of SU(3) to the SU(2) algebra which have structure constants of $f^{246} = 1/2$ and $\epsilon^{xyz} = 1$, respectively.
The inclusion of the $1/2$ factor makes it possible to reimagine both the  \{2-4-6\} subalgebra and ${\bf{S}}/2$ as belonging to different irreducible representations of a `1/4-quantized' SU(2) Lie algebra.

%\section{$1/4$-quantized SU(2) Lie algebra:  \{2-4-6\} subalgebra of SU(3)}
\section{$1/4$-quantized SU(2) Lie algebra}

To understand this algebra, we define the generators of the algebra as $T_{2,4,6}$ such that they satisfy $[ T_2, T_4 ] = \frac{i}{2} T_6$, in any representation.
Drawing an analogy to typical SU(2), $T_6$ is analogous to $\sigma^z/2$, and so we define the highest $T_6$ state $\ket{\tj}$ such that $T_6 \ket{\tj} = \tj \ket{\tj}$.
In addition, we define raising/lowering operators $T^{\pm}$, which raise and lower the $T_6$ eigenvalue by $\pm 1/2$ (as shown in Appendix \ref{app_chevalley_basis}).
In the same spirit as SU(2), the existence of a `lowest' state constrains the allowed values of $\tj$ to be $\tj = \ell / 4$, where $\ell \in \mathbb{R}$; we refer to $\tj$ as `physical-spin' labels.
In the more formal language of Lie algebras, the `physical-spin' label is related to the Dynkin labels, $\tj = \tilde{\lambda}_1/4$ (as shown in Appendix \ref{app_chevalley_basis}).
The $\tj=1/4$ representation is a two-dimensional representation with $T_2 = \frac{\sigma^x}{4}$, $T_4 =  \frac{\sigma^y}{4}$, $T_6 =  \frac{\sigma^z}{4}$, which is precisely ${\bf{S}}/2$. 
The $\tj=1/2$ representation is a three-dimensional representation with $T_2 = \frac{\lambda^2}{2}$, $T_4 =  \frac{\lambda^4}{2}$, $T_6 =  \frac{\lambda^6}{2}$ (this representation is used, up to unitary transformation, in the Kondo coupling).
Finally, the quadratic Casimir for a given $\tj$ representation is $Q = \tj ( \tj+1/2) \mathbb{I}$.

To understand the relation of the subalgebra to the typical SU(2) Lie algebra, we impose a generic normalization condition by fixing the length of the longest root $ | \alpha_1 ^2 | = \eta / 4,$ where $\eta \in \mathbb{R} >0$ is a chosen normalization convention.
Since the roots are the weights of the adjoint representation, the introduction of the $\eta$ normalization factor amounts to modification of the structure constants, and subsequently a redefinition of the generators. 
In particular, the redefined generators of SU(3), $\overline{T}_i \equiv \sqrt{\eta} {T}_i$, satisfy the Lie algebra $[\overline{T}_i, \overline{T}_j ] = \sqrt{\eta} f_{ij}^{k} \overline{T}_k $, where $T_i$ and $f_{ij}^{k}$ are the canonically normalized generators and structure constants of SU(3) used in constructing the Sugawara form.
This redefinition of the generators, results in a modification of the OPE of SU(3) currents,
%%%
\begin{align}
\overline{J}^a (z_1) \overline{J}^b (z_2) \sim \frac{1}{(z_{12})^2} \tilde{k} \delta^{ab} + \frac{1}{z_{12}} i \sqrt{\eta} f^{abc} \overline{J}^c
\end{align}
%%%
where $\tilde{k} \equiv \frac{\eta k}{2}$ and $z_{12} \equiv z_1 - z_2$, and the overline indicates current operators redefined to obey the generalized normalization.
We note that $k=2$ always.
We can now notice that for different choices of the normalization we can map the subalgebra to `more familiar' (canonically normalized) Lie algebras.
In the original $\eta = 1$ convention, we have the canonically normalized \{2-4-6\} subalgebra of SU(3).
For $\eta = 4$, the subalgebra gets mapped to the canonically normalized SU(2) Lie algebra. 
We can thus see that for particular choice of the normalization, we have a one-to-one correspondence between the original \{2-4-6\} subalgebra and SU(2).

With this generalized-normalization formulation, the affine extension of the subalgebra can thus be easily constructed in analogous methods to the conventional SU(2) case.
Due to the embedding of the \{2-4-6\} subalgebra in the larger affine SU(3)$_2$ algebra, the level of the embedded subalgebra is given by $\frac{8 \tilde{k}}{n} = 4k$.
This embedded level restricts the highest physical-spin state to be $\tj \leq k$.
As a consequence, the fusion coefficients can be similarly constructed and are of the form,
%%%
\begin{align}
\tj_1 \otimes \tj_2 = \bigoplus_{\tj_3} \mathcal{N}_{\tj_1, \tj_2} ^{(4k)} \ ^{\tj_3} 
\label{eq_fusion_coeff}
\end{align}
%%%
where 
\begin{align}
\mathcal{N}_{\tj_1, \tj_2} ^{(4k)} \ ^{\tj_3}  = 
\begin{cases}
    1,& \hspace{-2mm} \text{if } | \tj_1 - \tj_2| \leq \tj_3 \leq \text{min}(\tj_1 + \tj_2, \frac{2(4k)}{4} - \tj_1 - \tj_2) \\
    0,              & \text{otherwise}
\end{cases}
\end{align}
%%%
where we return back to the normalization of $\eta=1$.
\textcolor{black}{The fusion coefficients become an important ingredient in determining the modification of the free fermion boundary conditions due to the `absorption' of the multipolar impurity.}

\section{Maverick Coset Construction: \\ 3-state Potts model}

The full Hilbert space of the free electron model involves, U(1) charge, SU(3)$_2$ flavour, and SU(2)$_3$ spin degrees of freedom.
The symmetry group of the flavour degree of freedom is governed by $\mathcal{G} = SU(3)_2$, and involves currents $J_\mathcal{G}^a$; the subscript $\mathcal{G}$ is to explicitly recall that these currents belong to the complete SU(3)$_2$ group.
However, as seen in Eq. \ref{eq_select_generators}, only a select number of the SU(3)$_2$ bosonic currents couple to the local moment.
\textcolor{black}{This suggests that an alternative conformal embedding would be helpful when considering the the Kondo interaction.
To proceed,} we denote the subgroup $\mathcal{H} \in \mathcal{G}$, with currents $J_\mathcal{H}^a$ that belong to the \{2-4-6\} sector, and construct stress-energy tensors belonging to each of the groups,
%%%
\begin{align}
&T_\mathcal{G} (z) = \frac{1/2}{\tilde{k}+C_{A,\mathcal{G}}/2} \sum_{a=1}^8 : J_\mathcal{G}^a(z) J_\mathcal{G}^a(z): \\
&T_\mathcal{H} (z) = \frac{1/2}{\tilde{k}+C_{A,\mathcal{H}}/2} \sum_{a=1}^3 : J_\mathcal{H}^a(z) J_\mathcal{H}^a(z):
\end{align}
%%%
where $\tilde{k} = \eta k / 2$ just as before, and $C_{A}$ are the quadratic Casimirs of the adjoint representation of the respective groups i.e. $C_{A,\mathcal{G}} = 3 \eta$ and $C_{A,\mathcal{H}} = \eta / 2$.
The conformal weight/dimension of the primary states of a representation $\tj$ of $\mathcal{H}$ is given by,
%%%
\begin{align}
h_{\tj} = \frac{C_{\tj} / 2}{\tilde{k} + C_{A,\mathcal{H}}/2} = \frac{\frac{\eta}{2}\tj(\tj+1/2)}{\frac{\eta k }{2} + \frac{\eta}{4}} = \frac{\tj(\tj+1/2)}{k+1/2},
\end{align}
%%%
where in the final quality, we notice that normalization dependency of the quadratic Casimirs and $\tilde{k}$ drop out in the conformal weight/dimension.
Though the conformal weight is independent of the normalization, we notice that for $\eta=4$ (the normalization condition to map the \{2-4-6\} subalgebra to canonical SU(2)) the conformal weight of the primary states is identical to that of SU(2)$_8$, as expected.

The procedure of breaking up the stress-energy tensors is attractive.
This is because, though the currents $T_\mathcal{G}(z)$ and $T_\mathcal{H}(z)$ have singular contributions in their OPE with $J_\mathcal{H}^a$ individually, the OPE of $T_{\mathcal{G}/\mathcal{H}} \equiv T_\mathcal{G} - T_\mathcal{H}$ with $J_\mathcal{H}^a$ is non-singular, and subsequently so is the OPE of $T_{\mathcal{G}/\mathcal{H}}$ with all of $T_\mathcal{H}$ \cite{ginsparg_cft}.
As such, the Virasoro algebra generated by $T_\mathcal{G}$ can be decomposed into two mutually commuting Virasoro algebras, $[T_{\mathcal{G}/\mathcal{H}} ,  T_\mathcal{H}] = 0$.
This formulation, known as a coset construction, permits the efficient `breaking up' of a larger group into smaller subgroups.
To understand the nature of the coset, $\mathcal{G}/\mathcal{H}$, we consider its central charge,
%%%
\begin{align}
c_{\mathcal{G}/\mathcal{H}} = c_\mathcal{G} - c_\mathcal{H} = \frac{\tilde{k} \dim| \mathcal{G} |}{\tilde{k} + C_{A,\mathcal{G}}/2} -  \frac{\tilde{k} \dim| \mathcal{H} |}{\tilde{k} + C_{A,\mathcal{H} }/2} = \frac{4}{5}
\end{align}
%%%
where $\dim| \mathcal{G} | = 8$ and $\dim| \mathcal{H} | = 3$.
The central charge is independent of the normalization convention, as can be easily seen by both $\tilde{k}$ and $C_A$ scaling linearly in $\eta$.
The central charge of $4/5$ corresponds to the three-state Potts model, in addition to the minimal $\mathcal{M}(6,5)$ model.
Inspired by the normalization of $\eta=4$ resulting in mapping the \{2-4-6\} subalgebra to the canonically normalized SU(2) algebra of level $4k = 8$, we consider this coset model to be described by the maverick coset formulation SU(3)$_2$/SU(2)$_8$ \cite{maverick_coset_1, maverick_coset_2}.

Employing this coset formulation, the IR Hamiltonian in Eq. \ref{eq_sugawara_complete} can be reimagined as a U(1) $\times$ (3-state Potts model) $\times$ $\widetilde{\text{SU}}(2)_8$ $\times$ SU(2)$_3$ KM invariant conformal field theory; the $\sim$ overline denotes $1/4$-quantization of the SU(2)$_8$ KM algebra.
In order to make this association, it requires the so-called branching functions which act as \textcolor{black}{further} conditions to describe which irreps of the parent SU(3)$_2$ are associated with which primary field of the 3-state Potts model and which irrep of SU(2)$_8$.
We list the branching functions of this maverick coset model and the field associations in Appendix \ref{app_coset_branching_rules}.
We highlight that only four of the (3-state Potts model) primary fields enter into our identifications: ``identity'' field ($\mathbb{I}$), thermal operator field ($\epsilon$), spin operator field ($\sigma $), and $\mathcal{Z}$ field with corresponding conformal weight/dimension of $0$, $2/5$, $1/15$, and $2/3$, respectively.

\section{Finite size spectrum}

The above formulated theory is appropriately prepared for the application of boundary conformal field theory (B-CFT) of Cardy \cite{cardy_bcft_1, cardy_bcft_2} and an extension of the approach as applied by Affleck and Ludwig \cite{affleck_ludwig_overscreened, affleck_cft_review_1}.
The essence of Cardy's B-CFT \cite{cardy_bcft_1, cardy_bcft_2} is that it avoids handling of complicated boundary conditions, and instead focuses on boundary states; \textcolor{black}{Appendix \ref{app_cardy} briefly describes this relationship.}
These boundary states are related to the \textcolor{black}{multiplicity} coefficients $n_{AB}^\alpha$ which determine the spectrum of excitations for particular boundary conditions.
In our context, $\alpha = (Q, j_f, \text{[3-state Potts model field]}, \tj)$, where each label denotes the highest state of the respective algebra; and the non-trivial boundary condition (B) is due to having a single multipolar impurity at the origin. 
The corresponding non-trivial boundary states can be obtained from known (free-fermion, A) boundary states by the elegant ``fusion rule'' hypothesis of Affleck and Ludwig \cite{affleck_ludwig_overscreened, affleck_cft_review_1}, whereby the fusion coefficients of Eq. \ref{eq_fusion_coeff} relate the \textcolor{black}{multiplicity} coefficients of different boundary conditions.
Since, the impurity only couples to the $\widetilde{\text{su}}(2)_8$ irreps ($\tj$ quantum number), the fusion is performed within the $\widetilde{\text{su}}(2)_8$ sector only, leaving the other irrep labels unchanged i.e. $n^{\tj}_{AB} = \sum_{\tj} \mathcal{N}^{\tj} _{\tj_2, \tj_{I}} n^{\tj_2}_{AA}$, where we take $\tj_{I} = 1/4$ as the impurity spin belongs to the $\tj=1/4$ representation.

One of the key features of B-CFT is the existence of boundary operators with non-trivial conformal dimensions.
These operators give rise to singular contributions to the free energy, and subsequently to the various response functions of interest. 
Under the conformal mapping of a semi-infinite plane ($x-t$ half plane) with  particular choice of boundary conditions to a finite strip, the primary boundary operators with boundary condition $B$ on the plane are in a one-to-one correspondence to the states of the finite-strip with boundary conditions $BB$ (both ends of strip) \cite{affleck_cft_review_1}.
Most importantly, the energies of the states in the finite-strip conformal tower are the scaling dimensions of the boundary operators.
In our context, this amounts to considering the free-fermion tower, and performing a double fusion so as to obtain the appropriate scaling dimensions of the boundary operators.
Indeed it is the lowest-lying state (after the double fusion) that gives rise to the leading irrelevant boundary operator.

Employing the conformal embedding (which relates the the U(1) charge, $Q$, to the irreps of flavour, SU(3)$_2$, and spin, SU(2)$_3$) and the coset branching rules (which relates the irreps of SU(3)$_2$ to the coset fields and $\widetilde{\text{su}}(2)_8$ irreps), the energy of a primary state is given,
%%%
\begin{align}
\hspace{-2mm} E_{\text{tot}} = \frac{\pi}{l} \Big[	\frac{Q^2}{12} + \frac{j_f (j_f + 1)}{3+2} + h_{\text{coset}} + \frac{{\tj}({\tj}+1/2)}{2+ 1/2}		\Big],
\label{eq_finite_size_equation}
\end{align}
%%%
where $ h_{\text{coset}} $ is the scaling dimension of the coset primary field (Appendix \ref{app_coset_branching_rules}), $j_f$ labels the irrep of SU(2)$_3$ flavour degree of freedom, ${\tj}$ denotes the $1/4$-quantized SU(2)$_8$ degree of freedom.
Table \ref{tab_free_fermion} lists the finite-size energy spectrum for the free-fermion states, as given by Eq. \ref{eq_finite_size_equation}.
%%%%%%
\begin{table}[t]
\begin{tabular}{c|c|c|c|c}
\hline
$Q$, U(1) & $j_f$, SU(2)$_3$ & [3-state Potts model] & ${\tj}$, $\widetilde{\text{SU}}(2)_8$ & $\left(\frac{l}{\pi} \right)E_{\text{tot}}$  \\ \hline
0 & 0 & $\mathbb{I}$ & 0 & 0 \\ 
1 & $\frac{1}{2}$ & $\sigma$ & $\frac{1}{2}$ & $\frac{1}{2}$ \\ 
0 & 1 & $\mathbb{I}$ & 1 & 1 \\ 
0 & 1 & $\epsilon$ & $\frac{1}{2}$ & 1 \\ 
2 & 0 & $\sigma$ & 1 & 1 \\ 
2 & 0 & $\mathcal{Z}$ & 0 & 1 \\ 
2 & 1 & $\sigma$ & $\frac{1}{2}$ & 1 \\ 
\vdots & \vdots & \vdots & \vdots & \vdots
\end{tabular}
\caption{Free-fermion tower with anti-periodic boundary conditions at $x=l$. 
The employed primary fields of the 3-state Potts model are $\mathbb{I}$, $\epsilon$, $\sigma$, and $\mathcal{Z}$.
We note that we only consider the primary states here, and present states with energies $E_{\text{tot}} \leq \pi/l$; the remaining primary states are given in Appendix \ref{app_remaining_conformal_tower}.}
\label{tab_free_fermion}
\end{table}
%%%%%%

The impact of the multipolar impurity is accounted for by the double fusion ($n^{\tj}_{BB} = \sum_{\tj_2, \tj_3} \mathcal{N}^{\tj} _{\tj_2, \tj_{I}} \mathcal{N}^{\tj_2} _{\tj_3, \tj_{I}} n^{\tj_3}_{AA}$)  with the free-fermion $\tilde{j}$ states to yield the conformal tower in Table \ref{tab_double_fusion}.
\textcolor{black}{Under the double fusion, the ``spin'' $\tilde{j}$ label changes to: $0 \rightarrow 0, \frac{1}{2}$; $\frac{1}{2} \rightarrow 0, \frac{1}{2}, 1$, $1 \rightarrow \frac{1}{2}, 1, \frac{3}{2}$, etc.}
We emphasize that only the `re-shuffling' of the $1/4$-quantized SU(2)$_8$ states takes place, with the irreps of the other degrees of freedom remaining unaffected.
%%%%%%
\begin{table}[t]
\begin{tabular}{c|c|c|c|c}
\hline
$Q$, U(1) & $j_f$, SU(2)$_3$ & [3-state Potts model] & ${\tj}$, $\widetilde{\text{SU}}(2)_8$ & $\left(\frac{l}{\pi} \right)E_{\text{tot}}$  \\ \hline
%%%
0 & 0 & $\mathbb{I}$ & 0 & 0 \\ 
0 & 0 & $\mathbb{I}$ & $\frac{1}{2}$ & $\frac{1}{5}$ \\ 
%%%
1 & $\frac{1}{2}$ & $\sigma$ & 0 &  $\frac{3}{10}$ \\ 
1 & $\frac{1}{2}$ & $\sigma$ & $\frac{1}{2}$ &  $\frac{1}{2}$ \\ 
0 & 1 & $\mathbb{I}$ & $\frac{1}{2}$ & $\frac{3}{5}$   \\ 
2 & 0 & $\sigma$ & $\frac{1}{2}$ & $\frac{3}{5}$  \\ 
0 & 1 & $\epsilon$ & 0 & $\frac{4}{5}$ \\ 
2 & 1 & $\sigma$ &0 & $\frac{4}{5}$  \\ 
1 & $\frac{1}{2}$ & $\sigma$ & 1 &  $\frac{9}{10}$ \\ 
%%%
0 & 1 & $\mathbb{I}$ & 1 &  1  \\ 
0 & 1 & $\epsilon$ & $\frac{1}{2}$ & 1 \\ 
2 & 0 & $\mathcal{Z}$ & 0 & 1  \\ 
2 & 0 & $\sigma$ & 1 & 1 \\ 
2 & 1 & $\sigma$ & $\frac{1}{2}$ & 1 \\ 
\vdots & \vdots & \vdots & \vdots & \vdots
%%%
\end{tabular}
\caption{Primary operator content after double fusion of free-fermion tower in Table \ref{tab_free_fermion} with multipolar impurity at $x=0$ and $x=l$. 
The employed primary fields of the 3-state Potts model are $\mathbb{I}$, $\epsilon$, $\sigma$, and $\mathcal{Z}$.
Only the primary states of energies $E_{\text{tot}} \leq \pi/l$ are listed here; the remaining states are given in Appendix \ref{app_remaining_conformal_tower}.}
\label{tab_double_fusion}
\end{table}
%%%%%%
The lowest non-zero energy state corresponds to a primary state that is chargeless ($Q=0$), flavourless ($j_f=0$), coset-trivial ($\mathbb{I}$), and of $1/4$-quantized ``spin'' ${\tj}=1/2$.
The scaling dimension of the corresponding boundary operator, $\vec{\phi}$, is $\Delta = 1/5$.

Any operator that enters into the fixed point Hamiltonian must preserve the symmetry the U(1) $\times$ [3-state Potts model] $\times$ $\widetilde{\text{SU}}(2)_8$ $\times$ SU(2)$_3$ KM invariance of the conformal field theory.
On physical grounds, we also expect that the leading irrelevant operator will involve the primary ``spin'' field $\vec{\phi}$, as the Kondo coupling occurs between the multipolar impurity and the conduction electrons occurs in the $1/4$-quantized ``spin'' sector.
In order to meet the symmetry requirement, we can simply consider the application of a KM current density operator, $\mathcal{J}_{n<0} ^a$ (where $a = \{2,4,6\}$) which generates descendent states/operators when acted upon a primary state/operator.
Thus, the candidate for the leading irrelevant operator (that obeys the KM symmetry) is the first-descendent operator, $\vec{\mathcal{J}}_{-1} \cdot \vec{\phi}$, which is explicitly a ``spin-less'' object (as it is a scalar product of two ``spin'' operators), and has a scaling dimension of $1 + \Delta$.
\\
\section{Physical Properties: \\ Specific heat, resistivity, entropy}
\label{sec_physical_properties}

\textcolor{black}{
To characterize the novel fixed point, we first consider its impact on the thermodynamic and transport response functions such as specific heat and resistivity.
As detailed by Affleck and Ludwig \cite{affleck_ludwig_overscreened}, the leading irrelevant operator plays a central role in determining the scaling behaviour of these response functions.
In particular, the fixed point Hamiltonian, and subsequently free energy, is augmented by the leading irrelevant operator at the boundary ($\delta g_k \vec{\mathcal{J}}_{-1} \cdot \vec{\phi}$), from which the specific heat is computed from second order perturbation theory in $\delta g_k$. 
This yields the specific heat coefficient scaling as $C/T \sim T^{-1 + 2\Delta} = T^{-3/5}$.
Similarly, the leading order corrections to the scattering rate (due to the impurity) are computed by perturbing the one-electron Green's function linearly by the leading irrelevant operator \cite{affleck_cft_multi}.
The subsequent correction to the resistivity scales as $\rho \sim T^{\Delta} = T^{1/5}$. 
The coefficient of this correction is $\delta g_k$, and so the sign of the deviation is determined by whether the intermediate fixed point is approached from above ($\delta g_k>0$) or below ($\delta g_k<0$).
}

\textcolor{black}{
Finally, the residual entropy, $\mathcal{S}_{\text{imp}}(T)$, which provides a measure for the ground state-degeneracy of the novel fixed point, can be computed via the modular S-matrix \cite{affleck_cft_multi,affleck_ludwig_cft_bethe_compare} (as described in Appendix \ref{app_entropy}).
The entropy is found to be $\mathcal{S}_{\text{imp}}(T=0) = \ln \left(\frac{\sin(\pi/5)}{\sin(\pi/10)} \right) \approx 0.643$, which is just under twice the residual entropy of the two-channel Kondo model fixed point.
}

\section{Discussions}

In this work, we elucidated the nature of the novel non-Fermi liquid fixed point in the multipolar Kondo problem by employing non-abelian bosonization, current algebra, and boundary conformal field theory approaches.
A crucial finding of our work is that the most natural language to express the multipolar Kondo couplings is in the spin-orbital entangled $\ket{j, m_j}$ basis.
This suggests that though conduction electrons may have spin and orbital quantum numbers to begin with, their interaction with a single multipolar multipolar forces the two decoupled degrees of freedom to become intertwined.
This provides a route to controllably introduce spin-orbital entangling in a metallic system with the incorporation of multipolar impurities.

From the finite-size spectrum, the scaling behaviour of experimentally relevant quantities associated with the fixed point are obtained non-perturbatively; in particular, the specific heat coefficient $C/T \sim T^{-1 + 2\Delta} = T^{-3/5}$ and the resistivity $\rho \sim T^{\Delta} = T^{1/5}$.
\textcolor{black}{As well, the residual entropy, $\mathcal{S}_{\text{imp}}(T=0) \approx 0.643$, is almost twice as compared to the two-channel Kondo model \cite{affleck_ludwig_cft_bethe_compare}.}
The highly singular nature of the response functions concretely establishes the non-Fermi liquid nature of the novel fixed point.
We emphasize that though the scaling behaviour of the leading irrelevant operator looks the same as in the 8-channel dipolar Kondo model, the nature of the fixed point is different due to the other conformal sectors.
More specifically, the complete finite size spectrum is distinct in the novel fixed point, which can be tested by future numerical renormalization group computations.

\textcolor{black}{
An important requirement in realizing such an exotic Kondo effect is to ensure the symmetry-protected degeneracies are preserved.
Breaking the local $T_d$ symmetry (leading to splitting of the non-Kramers doublet ground state), or lifting the cubic symmetry of the underlying lattice (breaking the degeneracy of the conduction electron bands) by an external perturbation results in the demise of this Kondo problem, as the Kondo interactions describe energy preserving scattering events.
In effect, symmetry-breaking perturbations are relevant in the renormalization group sense for the non-trivial fixed point.
This can be understood from more general considerations, where one examines the influence of a field ($h_{\phi}$) conjugate to a boundary operator ($\phi$) with scaling dimension $\Delta_{\phi}$. The associated term added to the Lagrangian is $\delta \mathcal{L} (h_{\phi})  = h_\phi \int d \tau \phi( \tau)$, where the spatial integral is absent as the boundary operator is confined to $\bf{x}=0$.
Under the standard renormalization group procedure ($\tau \rightarrow \tau' = \tau /b$, where $b>1$), the conjugate field scales as $b^{1- \Delta_{\phi}}$. 
The field conjugate to the primary boundary operator is thus relevant in the renormalization group sense.
Physically, this conjugate field can be associated as lattice-stress fields that couple to the multipolar impurity \cite{cox_zawadowski_book} and split the non-Kramers degeneracy.
We note that a symmetry-preserving perturbation, such as hydrostatic pressure or chemical pressure by substitution of the transition metal in the Pr(TM)$_2$Al$_{20}$, acts as an irrelevant perturbation to the non-trivial fixed point.
}

The solvable problem tackled in this work provides the foundation for a diverse variety of Kondo effects and non-Fermi liquids.
In particular, though our work is motivated by the Pr(Ti,V)$_2$Al$_{20}$ family and as such focuses on $\mathcal{O}_{20}$, $\mathcal{O}_{22}$ quadrupolar and $\mathcal{T}_{xyz}$ octupolar local moments, this is just one out of a myriad of possibilities.
Indeed, there are examples of other $4f$ \cite{shiina_pr_os_sb, ceb6_hfm, ceb6_quadrupolar, thalmeier2019multipolar} and $5f$ \cite{Chandra_2002, Tripathi_2005, Kotliar_Haule_2009, Kotliar_Haule_2011, chandra_hastatic, chandra_hastatic} electrons subjected to a non-cubic crystalline electric fields, which gives rise to different possible combinations of supported multipolar moments.
In addition, conduction electrons may arrive with their own diversity in their orbital degrees of freedom being beyond the cubic $p$ considered in this work \cite{dhva_nagashima_2013, dhva_nagashima_2020}.
The combination of these two sources of diversity suggests that many different Kondo effects may occur, which could lead to a multitude of non-Fermi liquid behaviours.
This study thus opens a new route and territory for achieving and studying exotic Kondo effects and novel non-Fermi liquids.

\textcolor{black}{In terms of future work, a thorough classification of the possible non-Fermi liquids that may occur in multipolar based compounds would be an intriguing and impactful study.}
As well, the extension of the single-impurity problem to the corresponding multipolar Kondo lattice problem \cite{Qimiao_kondo_destruction, flint_mft_quadrupolar_kondo} and associated quantum critical phenomena are outstanding questions of future research.

\section*{Acknowledgements}
We thank Piers Coleman and Sungjay Lee for helpful discussions.
This work was supported by NSERC of Canada. Y.B.K. is supported by the Killam Research Fellowship of the Canada Council for the Arts.

%\vfill

\appendix
\section*{Appendix}

\section{Cubic harmonics to spin-orbital entangled basis}
\label{app_change_of_basis}

In order to rewrite the Kondo coupling in terms of spin-orbital entangled $j, m_j$ basis, we need to perform a double change of basis.
Firstly, the cubic harmonics are rewritten in terms of $L=1$ orbital angular momentum degrees of freedom,
%%%
\begin{align}
&\ket{p_x} = \frac{1}{\sqrt{2}}\Big(\ket{1, -1} - \ket{1, 1} \Big)_L \\
&\ket{p_y} = \frac{i}{\sqrt{2}}\Big(\ket{1,-1} + \ket{1, 1} \Big)_L \\
&\ket{p_z} = \ket{1, 0}_L
\end{align}
%%%
where we use the basis $\ket{L, m_L}$ for the angular momentum orbital degrees of freedom (the subscript $L$ is a perpetual remainder).
Secondly, the spin and orbital degrees of freedom are entangled by angular-momentum addition via Clebsch Gordon coefficients,
%%%
\begin{equation}
\begin{aligned}
& \ket*{1,1; \frac{1}{2}, \uparrow} = \ket*{\frac{3}{2}, \frac{3}{2}}_J \\
& \ket*{1,1; \frac{1}{2}, \downarrow} = \frac{1}{\sqrt{3}} \left(\ket*{\frac{3}{2}, \frac{1}{2}}_J + \sqrt{2} \ket*{\frac{1}{2}, \frac{1}{2}}_J \right) \\
& \ket*{1,0; \frac{1}{2}, \uparrow} = \frac{1}{\sqrt{3}} \left(\sqrt{2} \ket*{\frac{3}{2},  \frac{1}{2}}_J - \ket*{\frac{1}{2},  \frac{1}{2}}_J \right)   \\
& \ket*{1,0; \frac{1}{2}, \downarrow} =  \frac{1}{\sqrt{3}} \left(\sqrt{2} \ket*{\frac{3}{2},  \frac{-1}{2}}_J + \ket*{\frac{1}{2},  \frac{-1}{2}}_J \right)   \\
& \ket*{1,-1; \frac{1}{2}, \uparrow} = \frac{1}{\sqrt{3}} \left(\ket*{\frac{3}{2},  \frac{-1}{2}}_J - \sqrt{2}  \ket*{\frac{1}{2},  \frac{-1}{2}}_J \right) \\
& \ket*{1,-1; \frac{1}{2}, \downarrow} = \ket*{\frac{3}{2}, \frac{-3}{2}}_J
\end{aligned}
\end{equation}
%%%
where on the left-hand-side of the equality, we use the notation of $\ket{L,m_L; s, s^z}$, where $s$ is the conduction electron spin degree of freedom; 
on the right-hand-side of the equality we use the notation for the spin-orbital entangled basis $\ket{J,m_J}_J$, where the subscript is a perpetual remainder of the $J$ basis.
In the main text, we drop this subscript as the additional clarification is not required.

\section{Basis for two-channel Kondo fixed point}
\label{app_basis_two_channel}

Tuning the couplings to fixed point I, only the $j=3/2$ conduction electron states survive.
The two decoupled pseudospin-1/2 operators in the $\ket*{\frac{3}{2}, m_j}$ space are,
%%%
\begin{equation}
\begin{aligned}
\tau_A^x &= \frac{1}{2} \ket*{\fr{3}{2}, \fr{3}{2}}\bra*{\fr{3}{2}, -\fr{1}{2}}+ \fr{1}{2} \ket*{\fr{3}{2},-\fr{1}{2}}\bra*{\fr{3}{2},\fr{3}{2}},  \\
\tau_A^y &= \frac{ -i }{2} \ket*{\fr{3}{2}, -\fr{1}{2}}\bra*{\fr{3}{2}, \fr{3}{2}} + \frac{i}{2} \ket*{\fr{3}{2},\fr{3}{2}}\bra*{\fr{3}{2},-\fr{1}{2}},  \\
\tau_A ^z &= \frac{1}{2}  \ket*{\fr{3}{2}, -\fr{1}{2}}\bra*{\fr{3}{2}, -\fr{1}{2}} - \frac{1}{2}\ket*{\fr{3}{2}, \fr{3}{2}} \bra*{\fr{3}{2}, \fr{3}{2}}, 
\end{aligned}
\end{equation}
%%%
and its partner $\tau^{x,y,z}_B: \tau^{x,y,z}_A |_{ m_j \rightarrow -m_j}$,
%%%
\begin{equation}
\begin{aligned}
\tau_B^x &= \frac{1}{2} \ket*{\fr{3}{2}, -\fr{3}{2}}\bra*{\fr{3}{2}, \fr{1}{2}}+ \fr{1}{2} \ket*{\fr{3}{2},\fr{1}{2}}\bra*{\fr{3}{2}, -\fr{3}{2}},  \\
\tau_B^y &= \frac{ -i }{2} \ket*{\fr{3}{2}, \fr{1}{2}}\bra*{\fr{3}{2}, -\fr{3}{2}} + \frac{i}{2} \ket*{\fr{3}{2}, -\fr{3}{2}}\bra*{\fr{3}{2},\fr{1}{2}},  \\
\tau_B ^z &= \frac{1}{2}  \ket*{\fr{3}{2}, \fr{1}{2}}\bra*{\fr{3}{2}, \fr{1}{2}} - \frac{1}{2}\ket*{\fr{3}{2}, -\fr{3}{2}} \bra*{\fr{3}{2}, -\fr{3}{2}} .
\end{aligned}
\end{equation}
%%%

\section{Degeneracy of strongly-coupled fixed point}
\label{app_strong_coupling}

The strong-coupling limit provides a means to verify the consistency of the existence of the perturbatively obtained fixed point.
In the case of a single coupling constant of the original isotropic Kondo problem, this limit is unique.
However, in the case of multiple couplings (as in our model), there is an inherent ambiguity as each of the coupling constants can be taken to infinity independently.
To circumvent this issue, we take the reasonable and elegant approach of Nozieres and Blandin \cite{noz_fixed_point}, where one takes each of the couplings to infinity while fixing their ratio to be that at the fixed point.
This is the simplest extension of the strong-coupling limit of the isotropic Kondo model, while at the same time accounting for the properties of the non-trivial fixed point by fixing the ratio.

The ground state of the strongly-coupled novel fixed point is four-fold degenerate,
%%%
\begin{equation}
\begin{aligned}
\ket{\text{GS}_{1}} &= \frac{1}{\sqrt{3}} \Bigg[ \ket{\uparrow}_f  \ket*{\frac{1}{2}, \frac{-1}{2} } + \ket{\downarrow}_f   \left( \ket*{\frac{3}{2}, \fr{3}{2}} + i \ket*{\fr{3}{2}, \fr{-1}{2} } \right) \Bigg] \\
\ket{\text{GS}_{2}} &= \frac{1}{\sqrt{3}} \Bigg[ \ket{\uparrow}_f  \left( \ket*{\frac{3}{2}, \fr{3}{2}} - i \ket*{\fr{3}{2}, \fr{-1}{2} } \right)  + \ket{\downarrow}_f \ket*{\frac{1}{2}, \frac{-1}{2} }    \Bigg] \\
\ket{\text{GS}_{3}} &= \frac{1}{\sqrt{3}} \Bigg[ \ket{\uparrow}_f  \left( \ket*{\frac{3}{2}, \fr{-3}{2}} - i \ket*{\fr{3}{2}, \fr{1}{2} } \right)  - \ket{\downarrow}_f \ket*{\frac{1}{2}, \frac{1}{2} }    \Bigg] \\
\ket{\text{GS}_{4}} &= \frac{1}{\sqrt{3}} \Bigg[ \ket{\uparrow}_f  \ket*{\frac{1}{2}, \frac{1}{2} } - \ket{\downarrow}_f   \left( \ket*{\frac{3}{2}, \fr{-3}{2}} + i \ket*{\fr{3}{2}, \fr{1}{2} } \right) \Bigg] \\
\end{aligned}
\end{equation}
%%%
where $\ket{\text{GS}_{1,2}}$ and $\ket{\text{GS}_{3,4}}$ respectively belong to the (i) and (ii) decoupled SU(3) sectors defined in Sec. \ref{sec_novel_fp}.
$\ket{\text{GS}_{1}}$ ($\ket{\text{GS}_{2}}$) is related to $\ket{\text{GS}_{3}}$ ($\ket{\text{GS}_{4}}$) by time-reversal symmetry.
The instability of the strongly-coupled fixed point can be further highlighted by placing the impurity on a one-dimensional line (parallel to the $\hat{\bf{x}}$ direction) and allowing tunnelling of conduction electrons to the nearest neighbouring sites on either side of the impurity.
The tunnelling acts a perturbation to the strongly-coupled Kondo Hamiltonian.
We find that a two-fold degenerate ground state remains at the impurity site (with up to second nearest neighbour hopping), where one of the ground states is an equal superposition of $\ket{\text{GS}_{1,2}}$, while the other is an equal superposition of $\ket{\text{GS}_{3,4}}$.

\section{Operator product expansions of non-abelian currents}
\label{app_opes}

The operator product expansions (OPEs) of the non-abelian currents is crucial in determining the KM algebra as well as for rewriting the free-fermion Hamiltonian in terms of non-abelian bosonic currents (Sugawara form) \cite{ludwig_bosonize}.
The OPE for the U(1) charge, SU(3) current, and SU(2) currents are,
\begin{widetext}
%%%%
\begin{align}
& J(z_1) J(z_2)  = \frac{6}{z_{12}^2} +  : \psi^{\dag}_{L,\alpha,p} \psi_{L,\alpha,p}  \psi^{\dag}_{L,\beta,q} \psi_{L,\beta,q} : (z_2)  + \Big[	:(\partial_z \psi^{\dag}_{L,\alpha,p}) \psi_{L,\alpha,p} :  -  : \psi^{\dag}_{L,\alpha,p} (\partial_z \psi_{L,\alpha,p}) : \Big]  (z_2)  \\
& J^a(z_1) J^b(z_2)  = \frac{2}{2z_{12}^2} \delta^{ab} + \frac{i f^{abc}}{z_{12}}  J^c(z_2) + \frac{1}{4}   f^{abc} : \partial_z \left( \psi^{\dag}_{L,\alpha,p} \lambda^c _{\alpha \beta}  \psi_{L,\beta,p} \right): (z_2) +  \frac{1}{4} : \psi^{\dag}_{L,\alpha,p} \psi_{L,\beta,p}  \psi^{\dag}_{L,\gamma,q} \psi_{L,\delta,q} : (z_2) \lambda^a _{\alpha \beta} \lambda^b _{\gamma \delta}	\nonumber		\\
& ~~~~~~~~~~~~~~~~~~~ + \frac{1}{4} \left( \frac{2}{3} \delta^{ab}_{\alpha \beta}  + d^{abc} \lambda^c _{\alpha \beta}   \right)  \Big[	:(\partial_z \psi^{\dag}_{L,\alpha,p}) \psi_{L,\beta,p} :  -  : \psi^{\dag}_{L,\alpha,p} (\partial_z \psi_{L,\beta,p}) : \Big]  (z_2)  \\
%%%%%
& J^A(z_1) J^B(z_2)  = \frac{3}{2 z_{12}^2} \delta^{ab} + \frac{i \epsilon^{ABC}}{z_{12}}  J^C(z_2) + \frac{1}{4}   \epsilon^{ABC} : \partial_z \left( \psi^{\dag}_{L,\alpha,p} \sigma^C _{pq}  \psi_{L,\alpha,q} \right): (z_2) + \frac{1}{4} : \psi^{\dag}_{L,\alpha,p} \psi_{L,\alpha,q}  \psi^{\dag}_{L,\beta,s} \psi_{L,\beta,t} : (z_2) \sigma^A _{p q} \sigma^B _{s t}	\nonumber		\\
& ~~~~~~~~~~~~~~~~~~~ + \frac{1}{4} \left( \delta^{ab}_{p q}   \right)  \Big[	:(\partial_z \psi^{\dag}_{L,\alpha,p}) \psi_{L,\alpha,q} :  -  : \psi^{\dag}_{L,\alpha,p} (\partial_z \psi_{L,\alpha,q}) : \Big]  (z_2)
\end{align}
%%%%
\end{widetext}
%%%%%
where repeated indices $\{ \alpha, \beta, \gamma, \delta \}= \{1,2,3\}$, $c = \{1,2,3\}$, $C=\{x,y,z\}$, $\{p, q, s, t \}= \{1,2\}$ are summed over.
The level of each KM algebra can thus be read off directly from the numerator of the first term in each OPE.
We note that we use the canonical normalization of the structure constants to compute the OPEs.

\section{1/4-quantized SU(2) Lie algebra: Dynkin labels and ``physical-spin'' weight}
\label{app_chevalley_basis}

The 1/4-quantized SU(2) Lie algebra is analogous to SU(2), and as such we focus on some of the key differences (and draw analogies, when applicable) with SU(2).
As described in the main text, we define raising/lowering operators $T^{\pm} \equiv ({T_2 \pm i T_4})/\sqrt{2}$, which satisfy
%%%
\begin{align}
[T_6, T^{\pm}] = \pm \frac{1}{2} T^{\pm}, ~~~~~~~~~ [T^+, T^-] =  \frac{1}{2} T_6.
\label{eq_lie_algebra_orig}
\end{align}
%%%
The action of the these operators on the eigenstates of $T_6$ are,
%%%
\begin{align}
&T_6 \ket{\tj,m} &&= m \ket{\tj,m} \\
& T^- \ket{\tj,m} &&= N_m \ket{\tj, m-1/2} \\
& T^+ \ket{\tj,m-1/2} &&= N_m \ket{\tj,m}
\end{align}
%%%
where $N_m = \frac{1}{2 \sqrt{2} } \sqrt{ (2\tj + 2m)(2\tj -2m + 1) }$.
We can thus notice that similar to SU(2), there is one label for the highest-state ($\tilde{j}$), and one `ladder' to ascend and descend with $T^{\pm}$.

To apply the machinery of (affine) Lie algebra, it is helpful to recapture the above notation in a more formal setting.
In particular, a given representation is denoted by the highest weight/state, $\tilde{\lambda}$, which can written in the basis of fundamental weights, $\omega_1 = \alpha_1 / 2$, with integer coefficients ($\tilde{\lambda_1}$) known as Dynkin labels, $\tilde{\lambda} = \tilde{\lambda}_1 \omega_1$. 
The Dynkin labels are the eigenvalues of the Chevalley basis, and the relation between the Dynkin labels to the ``physical-spin'' weight is given by $\tj = \tilde{\lambda}_1/4$.
This can be seen by considering the typical Lie algebra bases.
In the Chevalley basis, the commutation relation is given by
%%%
\begin{align}
[e,f]=h, ~~~~~ [h,e] = 2e, ~~~~~ [h,f] = -2f
\end{align}
%%%
where the Cartan Matrix is the same as that of the canonical SU(2), $A = (2)$.
The eigenvalues of the Chevalley generator, $h$, are the Dynkin labels,
%%%
\begin{align}
h \ket{\tilde{\lambda}} = \tilde{\lambda}_1 \ket{\tilde{\lambda}}.
\end{align}
%%%
Taking a generic normalization factor (${\eta}$ just as in the main text) for the highest root, we have the corresponding generators in the Cartan-Weyl basis,
%%%
\begin{align}
H = \frac{h }{4}\sqrt{\eta}, ~~~~ E^+ = e ~~~~~ E^- = f,
\end{align}
%%%
which satisfy the commutation relations $[E^+, E^-] = \frac{4}{\sqrt{\eta}} H$ and $[H, E^{\pm}] = \pm\frac{\sqrt{\eta}}{2} E^{\pm}$.
Finally, we have the `physical' normalization-dependent generators, $\overline{T}^{\pm} = \sqrt{\frac{\eta}{8}} E^{\pm}$ and $\overline{T}_6 = H$, which satisfy the commutation relations in Eq. \ref{eq_lie_algebra_orig} with normalization dependency $\sqrt{\eta}$ on the right-hand-side.
Thus, the eigenvalue of $\overline{T}_6$ is given by, 
%%%%
\begin{align}
\overline{T}_6 \ket{\tilde{\lambda}} &= H \ket{\tilde{\lambda}}  = \frac{\sqrt{n}h}{4} \ket{\tilde{\lambda}}  = \frac{\sqrt{n}\tilde{\lambda}_1}{4} \ket{\tilde{\lambda}}. 
\end{align}
%%%
For $\eta=1$, we can notice that $\tj = \tilde{ \lambda}_1 / 4$, which gives us the interpretation of the \{2-4-6\} subalgebra as `1/4-quantized'.
Similarly, for $\eta=4$, the algebra satisfies the `canonically normalized' SU(2) Lie algebra, with the physical weight/label $\tilde{ \lambda}_1/2$.

\section{Conformal embedding: \\ SU(3)$_2 \oplus $ SU(2)$_3 \subset$ SU(6)$_1$}
\label{app_conformal_embedding}
Branching rules provide the decomposition coefficients of an irrep of a (affine) Lie algebra $g$ ($\hat{g}_k$) into the irreps of a smaller (affine) Lie algebra $p \subset g$ ($\hat{p}_{k'} \subset \hat{g}_k$).
In our context, we are interested in the decomposition of the irreps of SU(6)$_1$ into SU(2)$_3 \oplus$ SU(3)$_2$.
Since the level of SU(6) is 1, the only dominant highest-weight representations are the fundamental representations $\hat{\omega}_{0,1,\dots 5}$.
Following the procedure detailed in Chp. 17.A of Ref. \onlinecite{cft_book} of employing Young tableaux and outer automorphisms, we obtain the following decomposition in Eq. \ref{eq_yng_table},

%%%
\begin{equation}
     \Yvcentermath1
\Yboxdim{8pt}     
\begin{aligned}     
     \bullet & \mapsto \Bigg( \  \bullet \otimes \bullet \ \Bigg) \oplus \Bigg( \ \yng(2) \otimes \yng(2,1)  \ \Bigg) \\ \\
     \yng(1) & \mapsto \Bigg( \ \yng(1) \otimes \yng(1) \ \Bigg) \oplus \Bigg(  \ \yng(3) \otimes \yng(2,2) \  \Bigg) \\ \\
     \yng(1,1) & \mapsto \Bigg( \ \bullet \otimes \ \yng(2) \ \Bigg) \oplus \Bigg( \  \yng(2) \otimes \yng(1,1) \  \Bigg) \\ \\
     \yng(1,1,1) & \mapsto \Bigg( \ \yng(3) \otimes \bullet \ \Bigg) \oplus \Bigg(  \ \yng(1) \otimes \yng(2,1) \  \Bigg) \\ \\
     \yng(1,1,1,1) & \mapsto \Bigg( \ \bullet \otimes \  \yng(2,2) \ \Bigg) \oplus \Bigg(  \ \yng(2) \otimes \yng(1)  \ \Bigg) \\ \\
     \yng(1,1,1,1,1) & \mapsto \Bigg( \ \yng(1) \otimes \  \yng(1,1) \ \Bigg) \oplus \Bigg( \ \yng(3) \otimes \yng(2) \  \Bigg)
\end{aligned}
\label{eq_yng_table}
\end{equation}
%%%
where on the right-hand-side, the first (second) Young tableaux labels irreps of SU(2)$_3$ (SU(3)$_2$).
We can rewrite the above branching rules in terms of the Dynkin labels of the corresponding irreps, namely,
%%%
\begin{equation}
\begin{aligned}
& \hat{\omega}_0 \mapsto \Big( \left[ 3,0 \right]	\otimes \left[2,0,0 \right]  \Big) \oplus  \Big( \left[ 1,2 \right]	\otimes \left[0,1,1 \right]	\Big)  \\ 
& \hat{\omega}_1 \mapsto \Big( \left[ 2,1 \right]	\otimes \left[1,1,0 \right]  \Big) \oplus  \Big( \left[ 0,3 \right]	\otimes \left[0,0,2 \right]	\Big)  \\
& \hat{\omega}_2 \mapsto \Big( \left[ 3,0 \right]	\otimes \left[0,2,0 \right]  \Big) \oplus  \Big( \left[ 1,2 \right]	\otimes \left[1,0,1 \right]	\Big)  \\
& \hat{\omega}_3 \mapsto \Big( \left[ 0,3 \right]	\otimes \left[2,0,0 \right]  \Big) \oplus  \Big( \left[ 2,1 \right]	\otimes \left[0,1,1 \right]	\Big)  \\
& \hat{\omega}_4 \mapsto \Big( \left[ 3,0 \right]	\otimes \left[0,0,2 \right]  \Big) \oplus  \Big( \left[ 1,2 \right]	\otimes \left[1,1,0 \right]	\Big)  \\ 
& \hat{\omega}_5 \mapsto \Big( \left[ 2,1 \right]	\otimes \left[1,0,1 \right]  \Big) \oplus  \Big( \left[ 0,3 \right]	\otimes \left[0,2,0 \right]	\Big)  \\
\end{aligned}
\end{equation}
%%%
where we use the standard definition of the irreps of SU(N) in terms of Dynkin labels: $\hat{\lambda} = [\lambda_0, \lambda_1, \dots] = [k - \sum_{i}^{r} \lambda_i,  \lambda_1, \dots]$, where $r$ is the rank of the Lie algebra.

\section{Maverick coset branching rules}
\label{app_coset_branching_rules}

%%%
\begin{table*}[t]
\begin{tabular}{c | c | c | c | c | c}
$\hat{\lambda}$, SU(3)$_2$ & $\hat{\mu}$, SU(2)$_8$ & $ q^{-h_{\hat{\lambda}; \hat{\mu}}+c/24} \chi_{ \{ \hat{\lambda}; \hat{\mu} \} }$ & Grade of $\hat{\lambda}$ representation, $n$ & Conformal dimension, $h_{\hat{\lambda}; \hat{\mu}}$ & Field label  \\ 
\hline
$[2,0,0]$ & $[8,0]$ & \multirow{3}{*}{$1+q^2+2q^3+3q^4+4q^5+\dots $} & 0 &  \multirow{3}{*}{0}  & \multirow{3}{*}{$\mathbb{I}$} \\
$[2,0,0]$ & $[0,8]$ &  & 2  &  &  \\
$[0,1,1]$ & $[4,4]$ & ($= \chi_{1,1}^{\text{vir}} + \chi_{4,1}^{\text{vir}} $) & 0 &  &  \\
\hline
$[0,1,1]$ & $[6,2]$ & \multirow{3}{*}{$1+2q+2q^2+4q^3+5q^4+8q^5+\dots $} & 0 &  \multirow{3}{*}{$\frac{2}{5}$}  & \multirow{3}{*}{$\epsilon$} \\
$[0,1,1]$ & $[2,6]$ &  & 1  &  &  \\
$[2,0,0]$ & $[4,4]$ & ($= \chi_{2,1}^{\text{vir}} + \chi_{3,1}^{\text{vir}} $) & 1 &  &  \\
\hline
$[1,1,0]$ & $[6,2]$ & \multirow{6}{*}{$1+q+2q^2+3q^3+5q^4+7q^5+\dots $} & 0 &  \multirow{7}{*}{$\frac{1}{15}$}  & \multirow{7}{*}{$\sigma$} \\
$[1,1,0]$ & $[2,6]$ &  & 1  &  &  \\
$[0,0,2]$ & $[4,4]$ & $  $ & 0 &  &  \\
$[1,0,1]$ & $[6,2]$ & ($ =\chi_{2,3}^{\text{vir}}  $) & 0 &  &  \\
$[1,0,1]$ & $[2,6]$ & $  $ & 1 &  &  \\
$[0,2,0]$ & $[4,4]$ & $   $ & 0 &  &  \\
\hline
$[0,2,0]$ & $[8,0]$ & \multirow{6}{*}{$1+q+2q^2+2q^3+4q^4+5q^5+\dots $} & 0 &  \multirow{7}{*}{$\frac{2}{3}$}  & \multirow{7}{*}{$\mathcal{Z}$} \\
$[0,2,0]$ & $[0,8]$ &  & 2  &  &  \\
$[1,0,1]$ & $[4,4]$ & $  $ & 1 &  &  \\
$[0,0,2]$ & $[8,0]$ & ($ =\chi_{1,3}^{\text{vir}}  $) & 0 &  &  \\
$[0,0,2]$ & $[0,8]$ & $  $ & 2 &  &  \\
$[1,1,0]$ & $[4,4]$ & $   $ & 1 &  &  \\
\end{tabular}
%\end{ruledtabular}
\label{tab_coset_branchings}
\caption{Maverick coset SU(3)$_2 / $SU(2)$_8$ branching functions as computed in Refs. \onlinecite{maverick_coset_1, maverick_coset_2}. The grade denotes the descendent level of the irreps of SU(3)$_2$.
The minimal model field labelling is the same as that in Ref. \onlinecite{cft_book}.}
\end{table*}
%%%

The spirit of the coset branching rules is analogous to that of Appendix \ref{app_conformal_embedding} in that we once again are considering decompositions of the various representations, $\hat{\lambda}$, of the affine Lie algebra $\hat{g}$ into the representations, $\hat{\mu}$, of the affine Lie algebra $\hat{p}$, which is given by
%%%
\begin{align}
\hat{\lambda} \mapsto \bigoplus_{\hat{\mu}} b_{\hat{\lambda} \hat{\mu}} \hat{\mu}
\end{align}
%%%
where $b_{\hat{\lambda} \hat{\mu}}$ are the branching coefficients.
This decomposition can be rewritten in terms of characters,
%%%
\begin{align}
\chi_{\hat{\lambda} }  = \sum_{\hat{\mu}} \chi_{ \{ \hat{\lambda}; \hat{\mu} \} } \chi_{\hat{\mu} }
\end{align}
%%%
where $\chi_{\hat{\lambda}} \equiv q^{h_{\hat{\lambda}} - c/24} tr_{\hat{\lambda}}(q^{L_0})$. 
The term in the trace accounts for the `grade' or descendent level, $n$, of a state belonging to the $\hat{\lambda}$ irrep (i.e. already extracted out the conformal weight of the primary state, $h_{\hat{\lambda}}$).
For the maverick coset, the restricted character decompositions have been carefully computed in Refs. \onlinecite{maverick_coset_1, maverick_coset_2} , which we reproduce in Table \ref{tab_coset_branchings} with our notations.
Below each of the restricted characters in Table \ref{tab_coset_branchings}, we write down the combination of the 3-state Potts model characters (described below) that gives these restricted characters of the coset field.
This allows the the conformal dimension of the coset model to be obtained.
We note that the conformal dimension of the coset model is related to the conformal dimensions of the parent algebra ($\hat{g}$) and the subalgebra ($\hat{p}$): $h_{\chi_{ \{ \hat{\lambda}; \hat{\mu} \} }} = h_{\hat{\lambda}} +n - h_{\hat{\mu}} $; as such $n$ can be extracted. 
For reference, the characters of the 3-state Potts models \cite{cft_book} employed in Table \ref{tab_coset_branchings} are,
%%%
\begin{equation}
\begin{aligned}
& \chi_{1,1}^{\text{vir}} = q^{\frac{-1}{30}}\left(1+q^2+q^3+2q^4+2q^5+4q^6+\dots \right) \\
& \chi_{2,1}^{\text{vir}} = q^{\frac{2}{5} - \frac{1}{30}} \left( 1+ q + q^2 + 2q^3 + 3q^4 + 4q^5 + 6q^6 +\dots \right) \\
& \chi_{3,1}^{\text{vir}} = q^{\frac{7}{5} - \frac{1}{30}} \left( 1+ q + 2q^2 + 2q^3 + 4q^4 + 5q^5 + 8q^6 +\dots \right) \\
& \chi_{1,3}^{\text{vir}} = q^{\frac{2}{3} - \frac{1}{30}} \left( 1+ q + 2q^2 + 2q^3 + 4q^4 + 5q^5 + 8q^6 +\dots \right) \\
& \chi_{4,1}^{\text{vir}} = q^{3 - \frac{1}{30}} \left( 1+ q + 2q^2 + 3q^3 + 4q^4 + 5q^5 + 8q^6 +\dots \right) \\
& \chi_{2,3}^{\text{vir}} = q^{\frac{1}{15} - \frac{1}{30}} \left( 1+ q + 2q^2 + 3q^3 + 5q^4 + 7q^5 + 10q^6 +\dots \right)
\end{aligned}
\end{equation}
%%%

\section{Excited states of conformal towers}
\label{app_remaining_conformal_tower}
In the main text, we presented the conformal towers for energies $E_{\text{tot}} \leq \left(\frac{\pi}{l} \right)$.
We present in Table \ref{tab_free_fermion_higher_remain} and \ref{tab_double_fusion_remaining} the remaining, higher energy, states for the free-fermion and after the double-fusion with the impurity, respectively.
%%%%%%
\begin{table}[t]
\begin{tabular}{c|c|c|c|c}
\hline
$Q$, U(1) & $j_f$, SU(2)$_3$ & [3-state Potts model] & ${\tj}$, $\widetilde{\text{SU}}(2)_8$ & $\left(\frac{l}{\pi} \right)E_{\text{tot}}$  \\ \hline
\vdots & \vdots & \vdots & \vdots & \vdots \\
1 & $\frac{3}{2}$ & $\sigma$ & 1 & $\frac{3}{2}$ \\
1 & $\frac{3}{2}$ & $\mathcal{Z}$ & 0 & $\frac{3}{2}$ \\
3 & $\frac{3}{2}$ & $\mathbb{I}$ & 0 & $\frac{3}{2}$ \\
3 & $\frac{1}{2}$ & $\mathbb{I}$ & 1 & $\frac{3}{2}$ \\
3 & $\frac{1}{2}$ & $\epsilon$ & $\frac{1}{2}$ & $\frac{3}{2}$ \\
4 & 1 & $\sigma$ & $\frac{1}{2}$ & 2 \\
4 & 0 & $\sigma$ & 1 & 2 \\
4 & 0 & $\mathcal{Z}$ & 0 & 2 \\
5 & $\frac{1}{2}$ & $\sigma$ & $\frac{1}{2}$ & $\frac{5}{2}$ \\
5 & $\frac{3}{2}$ & $\sigma$ & 1 & $\frac{7}{2}$ \\
5 & $\frac{3}{2}$ & $\mathcal{Z}$ & 0 & $\frac{7}{2}$ \\
\end{tabular}
\caption{Free-fermion tower with anti-periodic boundary conditions at $x=l$. The primary states for $E_{\text{tot}} > \pi/l$ are presented here; primary states with $E_{\text{tot}} \leq \pi/l$ are given in Table \ref{tab_free_fermion}.}
\label{tab_free_fermion_higher_remain}
\end{table}
%%%%%%

%%%%%%
\begin{table}
\begin{tabular}{c|c|c|c|c}
\hline
$Q$, U(1) & $j_f$, SU(2)$_3$ & [3-state Potts model] & ${\tj}$, $\widetilde{\text{SU}}(2)_8$ & $\left(\frac{l}{\pi} \right)E_{\text{tot}}$  \\ \hline
\vdots & \vdots & \vdots & \vdots & \vdots \\
1 & $\frac{3}{2}$ & $\sigma$ & $\frac{1}{2}$ & $\frac{11}{10}$ \\
3 & $\frac{1}{2}$ & $\mathbb{I}$ & $\frac{1}{2}$ & $\frac{11}{10}$ \\
2 & 0 & $\mathcal{Z}$ & $\frac{1}{2}$ & $\frac{6}{5}$ \\ 
3 & $\frac{1}{2}$ & $\epsilon$ & 0 & $\frac{13}{10}$ \\
2 & 1 & $\sigma$ & 1 &  $\frac{7}{5}$ \\ 
0 & 1 & $\epsilon$ & 1 & $\frac{7}{5}$  \\ 
1 & $\frac{3}{2}$ & $\sigma$ & 1 &  $\frac{3}{2}$ \\
1 & $\frac{3}{2}$ & $\mathcal{Z}$ & 0 &  $\frac{3}{2}$ \\
3 & $\frac{3}{2}$ & $\mathbb{I}$ & 0 & $\frac{3}{2}$ \\
3 & $\frac{1}{2}$ & $\mathbb{I}$ & 1 & $\frac{3}{2}$ \\
3 & $\frac{1}{2}$ & $\epsilon$ & $\frac{1}{2}$ & $\frac{3}{2}$ \\
0 & 1 & $\mathbb{I}$ & $\frac{3}{2}$ & $\frac{8}{5}$  \\ 
2 & 0 & $\sigma$ & $\frac{3}{2}$ & $\frac{8}{5}$  \\ 
4 & 0 & $\sigma$ & $\frac{1}{2}$ & $\frac{8}{5}$  \\
1 & $\frac{3}{2}$ & $\mathcal{Z}$ & $\frac{1}{2}$ &  $\frac{17}{10}$ \\
3 & $\frac{3}{2}$ & $\mathbb{I}$ & $\frac{1}{2}$ &  $\frac{17}{10}$ \\
4 & 1 & $\sigma$ & 0 & $\frac{9}{5}$ \\
3 & $\frac{1}{2}$ & $\epsilon$ & 1 & $\frac{19}{10}$  \\
%%%%
4 & 0 & $\sigma$ & 1 & 2 \\
4 & 1 & $\sigma$ & $\frac{1}{2}$ & 2 \\
4 & 0 & $\mathcal{Z}$ & 0 & 2  \\
1 & $\frac{3}{2}$ & $\sigma$ & $\frac{3}{2}$ & $\frac{21}{10}$ \\
3 & $\frac{1}{2}$ & $\mathbb{I}$ & $\frac{3}{2}$ & $\frac{21}{10}$ \\
4 & 0 & $\mathcal{Z}$ & $\frac{1}{2}$ & $\frac{11}{5}$ \\
5 & $\frac{1}{2}$ & $\sigma$ & 0 & $\frac{23}{10}$ \\
4 & 1 & $\sigma$ & 1 & $\frac{12}{5}$ \\
5 & $\frac{1}{2}$ & $\sigma$ & $\frac{1}{2}$ & $\frac{5}{2}$ \\
4 & 0 & $\sigma$ & $\frac{3}{2}$ & $\frac{13}{5}$ \\
5 & $\frac{1}{2}$ & $\sigma$ & 1 & $\frac{29}{10}$ \\
%%%%
5 & $\frac{3}{2}$ & $\sigma$ & $\frac{1}{2}$ & $\frac{31}{10}$ \\
5 & $\frac{3}{2}$ & $\sigma$ & 1 & $\frac{7}{2}$ \\
5 & $\frac{3}{2}$ & $\mathcal{Z}$ & 0 &  $\frac{7}{2}$ \\
5 & $\frac{3}{2}$ & $\mathcal{Z}$ & $\frac{1}{2}$ & $\frac{37}{10}$ \\
5 & $\frac{3}{2}$ & $\sigma$ & $\frac{3}{2}$ & $\frac{41}{10}$ \\
\end{tabular}
\caption{Primary operator content after double fusion of primary states of energies $E_{\text{tot}} > \pi/l$; the primary states with $E_{\text{tot}} \leq \pi/l$ are given in Table \ref{tab_double_fusion}.}
\label{tab_double_fusion_remaining}
\end{table}
%%%%%%

\section{Boundary conformal field theory and Residual Entropy}
\label{app_cardy}
\color{black}{
In this section, we briefly describe the rationale behind Cardy's boundary conformal field theory.
This discussion is employed in calculating the residual entropy in Appendix \ref{app_entropy}.
We direct the reader to the thorough references \cite{affleck_cft_review_1, cardy_bcft_1, cardy_bcft_2, cft_book} for detailed discussions.

As described in the main text, the impurity acts as a conformally invariant boundary condition to the bulk conformal theory.
In the spirit of Cardy \cite{cardy_bcft_1, cardy_bcft_2}, it is helpful to the consider the conformal system on a cylinder of spatial length $l$ and periodic temporal length $\beta$ that winds around the circumference.
The partition function for this system is then given by,
%%%
\begin{align}
\mathcal{Z}_{AB} = \text{tr} \left[ \exp( - \beta H_{AB} ) \right] = \sum_a n^{a}_{AB} \chi_a \left( e^ {-\frac{ \pi \beta }{l} } \right)
\label{eq_partition_thermal}
\end{align}
%%%
where the subscript on the Hamiltonian indicates the boundary conditions in the spatial direction, $n_{AB}^a$ are the \textcolor{black}{multiplicity} coefficients described in the main text, and $\chi_a \left( e^ {-\frac{ \pi \beta }{l} } \right)$ are the characters of the Virasoro algebra for the conformal tower $a$.
Equation \ref{eq_partition_thermal} can be understood as the quantum mechanical partition function computed over the `thermally coherent' time $\beta = 1/T$.

However, one can perform a modular transformation, which physically interchanges the temporal and spatial axes $\tau \leftrightarrow x$.
The partition function for this system is,
%%%
\begin{align}
\mathcal{Z}_{AB} ^{\tau \leftrightarrow x} = \bra{A} \exp( - l H_{S}  ) \ket{B} = \sum_a \langle{A} | a \rangle \langle{a} | B \rangle  \chi_a \left( e^ {-\frac{4 \pi l }{\beta} } \right),
\label{eq_partition_propogator}
\end{align}
%%%
where the subscript $S$ indicates the modular transformation has been made, $\ket{a}$ symbolically denotes the so-called ``Ishibashi'' states that are used to enforce no momentum/energy flow across the boundary.
Due to the interchange of the temporal and spatial axes, Eq. \ref{eq_partition_propogator} can be understood as the propagation of the evolution operator over `time' $l$ between initial and final states $A$ and $B$.
The Virasoro characters in Eq. \ref{eq_partition_thermal} and \ref{eq_partition_propogator} are related by the modular S-matrix, which allows one to find a relationship between the boundary conditions (encoded in the \textcolor{black}{multiplicity} coefficients $n_{AB}^a$) and the boundary states ($\ket{A,B}$),
%%%
\begin{align}
\sum_b S^{a}_{b} n_{AB}^b = \langle{A} | a \rangle \langle{a} | B \rangle ,
\label{eq_cardy_equations}
\end{align}
%%%
where $S^{a}_{b}$ is the modular S-matrix, and $b$ sums over the conformal towers.
Equation \ref{eq_cardy_equations} is known as ``Cardy's equations'' that allow boundary conditions (left-hand side) to be related to boundary states (right-hand side) \cite{affleck_cft_review_1}.

\section{Residual entropy}
\label{app_entropy}

The impurity entropy is defined as \cite{affleck_cft_review_1},
%%%
\begin{align}
\mathcal{S}_{\text{imp}}(T) \equiv \lim_{l \rightarrow \infty} \left[ \mathcal{S}(l,T) - \mathcal{S}_0(l,T) \right],
\end{align}
%%%
where $\mathcal{S}_0 (l,T)$ is the free fermion entropy, and the large $l$ limit indicates the thermodynamic (macroscopic) limit; the zero temperature limit is taken after the macroscopic limit.
The large $l / \beta$ limit suggests that Eq. \ref{eq_partition_propogator} would be convenient to employ for the partition function, as only the lowest/ground state would need to be used from the summation.
The impurity entropy at $T=0$ can then be shown to be $\mathcal{S}_{\text{imp}}(T=0) = \ln  \langle{A} | 0 \rangle \langle{0} | B \rangle $, where $\ket{0}$ denotes the ground state.
This can be understood as the sum of entropies arising form the two boundaries $A,B$ \cite{affleck_cft_multi,affleck_ludwig_cft_bethe_compare}.
Applying the ``fusion rule'' hypothesis of Affleck and Ludwig along with Cardy's equations (in conjunction with the Verlinde Formula \cite{affleck_cft_review_1}), leads to the below Eq. \ref{eq_entropy}.
%$\mathcal{S}_{\text{imp}}(T=0) = \ln \left( {S^0_{\frac{1}{2}}} / {S^0_{0}} \right)$.
%
Considering our maverick coset formulation, we focus only on the $\widetilde{\text{SU}}(2)_8$ sector (taking $\eta = 4$ canonical normalization) as the remaining sectors are decoupled from the impurity.
This allows us to employ the modular S-matrix for canonically normalized SU(2)$_k$,
%%%
\begin{align}
S^{m}_{n} (k) = \sqrt{\frac{2}{k+2}} \sin \left[ \frac{\pi (2m+1)(2n+1) }{k+2} \right].
\end{align}
%%%
The residual entropy (for generalized $k$ channels) is thus,
%%%
\begin{align}
\mathcal{S}_{\text{imp}}(T=0) = \ln \left( \frac{S^0_{1/2} (k) }{S^0_{0} (k) } \right) = \ln \left( \frac{\sin  \frac{2 \pi}{k+2} }{\sin\frac{\pi}{k+2}} \right).
\label{eq_entropy}
\end{align}
%%%
Taking $k=8$ in Eq. \ref{eq_entropy} yields the residual entropy given in Sec. \ref{sec_physical_properties}.
}

\color{black}

%\vfill
\clearpage
%bibliography

%\bibliography{cft_references_bibtex_response}

\end{document}